\documentclass[nofootinbib,twocolumn,floatfix,prd,preprintnumbers]{revtex4}
\usepackage{graphicx}
\usepackage{amsfonts}
\usepackage{amssymb}
\usepackage{amsmath}
\usepackage{bm}
\usepackage{dcolumn}
\usepackage{subfigure}
\usepackage{hyperref}

\begin{document}

\preprint{hep-th/0612202}

\title{Coupled bulk and brane fields about a de Sitter brane}

\author{Antonio Cardoso}
\email{antonio.cardoso@port.ac.uk}
\author{Kazuya Koyama}
\email{kazuya.koyama@port.ac.uk}
\author{Andrew Mennim}
\email{andrew.mennim@port.ac.uk}
\author{Sanjeev S.~Seahra}
\email{sanjeev.seahra@port.ac.uk}
\author{David Wands}
\email{david.wands@port.ac.uk}

\affiliation{Institute of Cosmology \& Gravitation, University of
Portsmouth, Portsmouth~PO1~2EG, UK}

\setlength\arraycolsep{2pt}

\newcommand*{\Y}{{Y}_{lm}}

\newcommand*{\Scalar}[1]{ {\mathsf{H}}_{#1} }
\newcommand*{\Tensor}{ {\mathsf{K}} }

\newcommand*{\dlangle}{\bm\{}
\newcommand*{\drangle}{\bm\}}
\newcommand*{\arcsinh}{\text{arcsinh}}

\newcommand*{\di}{\partial}
\newcommand*{\dihat}{\hat\partial}
\newcommand*{\dimT}{\Theta}
\newcommand*{\bulk}{\text{bulk}}
\newcommand*{\KG}{\text{\tiny KG}}
\renewcommand{\b}{\text{b}}
\renewcommand{\H}{\text{H}}
\newcommand{\J}{\text{J}}
\newcommand*{\rhohat}{\hat{\rho}}

\newcommand{\onehalf}{1/2}

\renewcommand{\Im}{\text{Im}\,}
\renewcommand{\Re}{\text{Re}\,}

\newcommand*{\q}{\tilde{q}}
\newcommand*{\phitilde}{{\tilde{\phi}}}
\newcommand*{\phitildeR}{{\tilde{\phi}^\text{\tiny R}}}
\newcommand*{\phiR}{\phi^\text{\tiny R}}
\newcommand*{\betacrit}{\beta_\text{crit}}
\newcommand*{\omegaphys}{\omega_\text{phys}}
\renewcommand*{\t}{\hat{t}}
\newcommand*{\z}{\hat{z}}
\newcommand*{\etahat}{\hat{\eta}}
\newcommand{\eff}{\text{eff}}
\newcommand{\betafour}{{}^{\text{\tiny (4) }}\!\!\beta}

\date{December 19, 2006}

\begin{abstract}

We consider the evolution of a bulk scalar field in anti-de Sitter
(AdS) spacetime linearly coupled to a scalar field on a de Sitter
boundary brane.  We present results of a spectral analysis of the
system, and find that the model can exhibit both bound and continuum
resonant modes. We find that zero, one, or two bound states may
exist, depending upon the masses of the brane and bulk fields
relative to the Hubble length and the AdS curvature scale and the
coupling strength. In all cases, we find a critical coupling above
which there exists a tachyonic bound state. We show how the
5-dimensional spectral results can be interpreted in terms of a
4-dimensional effective theory for the bound states. We find
excellent agreement between our analytic results and the results of
a new numerical code developed to model the evolution of bulk fields
coupled to degrees of freedom on a moving brane. This code can be
used to model the behaviour of braneworld cosmological perturbations
in scenarios for which no analytic results are known.

\end{abstract}


\maketitle

\section{Introduction}

Branes offer an alternative to the traditional Kaluza-Klein
approach to dimensional reduction. Some degrees of freedom are
restricted to branes, while other fields, including gravity,
propagate in the higher-dimensional bulk spacetime. In this case
the extra dimensions may be non-compact \cite{Randall:1999vf}.
Brane and bulk fields are usually treated as independent degrees
of freedom, but in general one should include interactions between
brane and bulk fields.  For instance, in order to study
inhomogeneous perturbations about a brane-world cosmology we need
to consistently include bulk metric perturbations coupled to
matter perturbations on the brane
\cite{Mukohyama:2000ui,Kodama:2000fa,vandeBruck:2000ju,Koyama:2000cc,%
Mukohyama:2000ga,Langlois:2000ph,Bridgman:2001mc}. Once we include
a finite coupling between brane and bulk fields it is no longer
possible to think in terms of a brane field without consistently
including the associated bulk field configuration, and vice versa.

In this paper we will consider the case of linear coupling between
a four-dimensional scalar field on a de Sitter brane and a
five-dimensional scalar field in an anti-de Sitter (AdS) bulk
spacetime. We will develop an analytic approach based on the
resonant scattering modes of the bulk field as has recently been
applied to the study of quasinormal modes of the gravitational
field about Randall-Sundrum \cite{Seahra:2005wk} and
Einstein-static \cite{Seahra:2005us,Clarkson:2005mg} branes.  We
identify both evanescent scattering modes (quasinormal modes) of
the bulk field, which decay at future null infinity, and
normalisable bound states. Our linearly coupled brane and bulk
perturbations about a de Sitter brane may be seen as a simplified
model of perturbations in the inflaton field driving slow-roll
inflation on a brane which are linearly coupled to scalar metric
perturbations in the bulk
\cite{Koyama:2004ap,Koyama:2005ek,Hiramatsu:2006cv}.

Linearly coupled brane and bulk fields on a Minkowski brane have
been previously studied in a Minkowski bulk
\cite{George:2004wh,Koyama:2005gh} and in an anti-de Sitter (AdS)
bulk \cite{Koyama:2005gh}. In a Minkowksi bulk there is always one
and only one bound state, and above a critical coupling the bound
state describes a tachyonic instability \cite{George:2004wh}. In
an AdS bulk one finds again a tachyonic bound state above a
critical coupling, but below the critical coupling there is no
bound state \cite{Koyama:2005gh}. On the other hand we know there
is a bound state for a light bulk field in the case of zero
coupling in the presence of a de Sitter brane. The spectrum of
free bulk modes is qualitatively different in the presence of a de
Sitter brane as only modes above a finite mass gap (proportional
to the Hubble rate on the brane) can escape to null infinity
\cite{Garriga:1999bq}. We will investigate the existence of bound
states with finite coupling between bulk and brane fields in the
presence of a de Sitter brane, finding zero, one or two bound
states depending upon the parameter values.

We also present a numerical code to solve for the evolution of the
bulk field coupled to the scalar field on a moving brane in the
bulk spacetime. This is a development of Seahra's numerical code
\cite{Seahra:2006tm} (see also
\cite{Hiramatsu:2006cv,Hiramatsu:2003iz,Ichiki:2003hf,Ichiki:2004sx,Hiramatsu:2004aa})
for bulk perturbations with an arbitrary brane trajectory but no
brane coupling, to include coupled degrees of freedom on the
brane. We use this to test our analytic results, but we also
expect this numerical code to be useful to investigate a wider
range of models such as metric perturbations in a bulk spacetime
which have a non-trivial coupling to the brane fields and are thus
less amenable to analytic study
\cite{Koyama:2004ap,Koyama:2005ek,Hiramatsu:2006cv}.

The layout of the paper is as follows:  In Sec.~\ref{sec:problem},
we introduce our model and the notation used throughout the paper.
Sec.~\ref{sec:spectral} discusses the spectral solution of the
equations of motion and the system's resonant excitations. An
effective 4-dimensional description of the model's bound state
sector is developed and compared to the full theory in
Sec.~\ref{sec:effective theory}.  Direct solution of the wave
equations in the time domain via numeric computation is the topic of
Sec.~\ref{sec:numeric}.  Finally, Sec.~\ref{sec:conclusions} is
reserved for our conclusions.

\section{Statement of the problem}\label{sec:problem}

In this section, we describe the model analyzed in this paper. We
consider a portion of AdS space $\mathcal{M}$ bounded by a brane
$\di\mathcal{M}_\b$. The bulk geometry can be usefully described by
two different coordinate systems
\begin{eqnarray}\nonumber
    ds_5^2 & = & \frac{\ell^2}{z^2} (-dt^2 + d\mathbf{x}^2 + dz^2) \\
    & = & \left( \frac{H\ell}{\sinh Hy} \right)^2 (-d\tau^2 + e^{2H\tau}
    d\mathbf{x}^2 + dy^2).
\end{eqnarray}
The first set of coordinates (``Poincar\'e'') are related to the
second set (``brane normal'') by
\begin{equation}\label{eq:coordinate transform}
    Ht = -e^{-H\tau}\cosh Hy, \quad Hz =
    e^{-H\tau}\sinh Hy.
\end{equation}
Here $H$ is the Hubble parameter: a positive constant with
dimensions $(\text{length})^{-1}$.

The trajectory of the brane in the brane normal coordinates is
simply
\begin{equation}
    y = y_\b = H^{-1}\arcsinh \, H\ell,
\end{equation}
which gives the induced line element
\begin{equation}\label{eq:brane metrics}
    ds_\b^2 = -d\tau^2 +e^{2H\tau} d\mathbf{x}^2 = (H\eta)^{-2}(-d\eta^2 +
    d\mathbf{x}^2),
\end{equation}
where we have defined the conformal time $\eta$ by
\begin{equation}\label{eq:eta}
    H\eta = -e^{-H\tau}, \quad \eta < 0.
\end{equation}
From (\ref{eq:brane metrics}), we see that $\tau$ is the proper
time on the brane.  We can also write down the trajectory of the
brane in Poincar\'e coordinates:
\begin{equation}\label{eq:trajectory}
    t = t_\b = \gamma \eta, \quad z = z_\b = -\eta
    H\ell,
\end{equation}
where
\begin{equation}
    \gamma = \sqrt{1+H^2\ell^2} = \cosh Hy_\b.
\end{equation}
We write the normal vector to the brane and the brane induced
metric as
\begin{subequations}
\begin{align}
    n^a\di_a & = \di_y = (-H\eta)(-H\ell\,\di_t+\gamma\,\di_z), \\
    h_{ab} & = g_{ab} - n_a n_b.
\end{align}
\end{subequations}
Finally, note that our bulk manifold $\mathcal{M}$ corresponds to
$y > y_\b$, which implies that $t < 0$ and $z
> 0$ by the coordinate transformation (\ref{eq:coordinate
transform}).  We assume the $\mathbb{Z}_2$ symmetry, which implies
that the ``other'' half of the bulk ($y \ngtr y_\b$) is the mirror
image of $\mathcal{M}$.

We consider two massive test scalar fields propagating on this
background geometry.  One of the fields $\phi$ exists throughout the
bulk, while the other $q$ is confined to the brane.  These are
governed by the action
\begin{multline}\label{eq:action}
    S = -\int_\mathcal{M} \left( g^{ab} \, \di_a \phi \, \di_b \phi + m^2
    \phi^2 \right) \\ - \frac{1}{2} \int_{\di\mathcal{M}_\b} \left( h^{ab}
    \, \di_a q \, \di_b q + \mu^2
    q^2  + 2\beta \phi q \right).
\end{multline}
The bulk part of the action is twice the usual expression due to
the $\mathbb{Z}_2$ symmetry.  The coupling constant $\beta$ has
dimensions of $(\text{length})^{-3/2}$ and can be taken to be
positive without loss of generality.  The masses of $\phi$ and $q$
are $m$ and $\mu$, respectively. Variation with respect to each
field yields the equations of motion
\begin{subequations}\label{eq:EOMs 1}
\begin{eqnarray}
    g^{ab}\nabla_a \nabla_b \phi & = & m^2\phi, \\
    h^{\alpha\beta}\nabla_\alpha \nabla_\beta q & = & \mu^2 q +
    \beta\phi_\b,
\end{eqnarray}
\end{subequations}
where $h_{\alpha\beta}$ and $\nabla_\alpha$ are the intrinsic metric
and covariant derivative on $\di\mathcal{M}_\b$, respectively, and
$\phi_\b$ is the bulk field evaluated on the brane. The bulk field
satisfies the boundary condition
\begin{equation}\label{eq:boundary 1}
    ( n^a \di_a \phi )_\b = \tfrac{1}{2}\beta q.
\end{equation}
In what follows, we will almost always work with the spatial Fourier
transforms of the fields
\begin{subequations}\label{eq:Fourier transform}
\begin{eqnarray}
    \phi(\tau,\mathbf{x},y) & = & \int d^3 k \,
    e^{-i\mathbf{k}\cdot\mathbf{x}} \phi_\mathbf{k} (\tau,y), \\
    q(\tau,\mathbf{x}) & = & \int d^3 k \,
    e^{-i\mathbf{k}\cdot\mathbf{x}} q_\mathbf{k} (\tau).
\end{eqnarray}
\end{subequations}
The $\mathbf{k}$ subscript on $\phi$ and $q$ will be omitted in all
the formulae below, with the exception of those in
Sec.~\ref{sec:effective theory}, where we work with the fields in
real space.

\section{Spectral analysis}\label{sec:spectral}

In this section, we seek the solution to the equations of motion
for our system (\ref{eq:EOMs 1}) via a spectral decomposition of
the fields. Our main goal will be to find and classify the
resonant modes in the problem by looking at the poles of the
scattering matrix.  Throughout this section, we use the brane
normal $(\tau,y)$ coordinates.

\subsection{Canonical wave equations and the scattering
matrix}\label{sec:scattering matrix}

To put the wave equations into canonical form we introduce the
re-scaled fields $\phitilde$ and $\q$, defined by
\begin{subequations}
\begin{eqnarray}
    \phi(\tau,y) & = & e^{-3H\tau/2}(\sinh Hy)^{3/2}
    \phitilde(\tau,y), \\
    q(\tau) & = & e^{-3H\tau/2} \q(\tau).
\end{eqnarray}
\end{subequations}
Then, the wave equations (\ref{eq:EOMs 1}) reduce to
\begin{subequations}\label{eq:EOMs 2}
\begin{align}
    \left[ \di_\tau^2 + k^2 e^{-2H\tau} \right] \phitilde & =
    \left[ \di_y^2 - U(y)
    \right] \phitilde, \\ \left[ \di_\tau^2 + k^2 e^{-2H\tau} \right]
    \q & = \left[ \tfrac{9}{4}H^2-\mu^2 \right] \q - \beta (H\ell)^{3/2} \phitilde_\b,
\end{align}
\end{subequations}
where we have defined the potential
\begin{equation}
    U(y) = \frac{H^2(15 + 4m^2\ell^2)}{4(\sinh Hy)^2}.
\end{equation}
The structure of the equations of motion (\ref{eq:EOMs 2}) suggest
a spectral decomposition of the form
\begin{subequations}\label{eq:decomposition 1}
\begin{eqnarray}
    \phitilde(\tau,y) & = & \int_{-\infty}^{\infty} d\omega \,
    T_\omega(\tau) \phitilde_\omega (y), \\
    \q(\tau) & = & \int_{-\infty}^{\infty} d\omega \,
    T_\omega(\tau) \q_\omega.
\end{eqnarray}
\end{subequations}
Here, $T_\omega(\tau)$ is an eigenfunction of the temporal operator
appearing in (\ref{eq:EOMs 2}):
\begin{equation}\label{eq:temporal eigenvalue}
    \left[ \di_\tau^2 + k^2 e^{-2H\tau} \right] T_\omega(\tau) =
    -\omega^2 H^2 T_\omega(\tau).
\end{equation}
This is satisfied if we take
\begin{equation}\label{eq:normalization}
    T_\omega(\tau) = \left(\frac{k}{2H} \right)^{i\omega} \left(
    \frac{H\omega}{2\sinh \pi\omega} \right)^{1/2} \J_{-i\omega} \left(
    \frac{k}{H}e^{-H\tau}\right),
\end{equation}
where $\J_\nu(x)$ is the Bessel function of the first kind. We note
that $T_\omega$ reduces to a single-frequency plane wave in the
limit of large $\tau$,
\begin{equation}
    T_\omega(\tau) \sim  \left(
    \frac{H\omega}{2\sinh \pi\omega} \right)^{1/2} \frac{e^{i\omega H\tau}}
    {\Gamma(1-i\omega)}, \quad [H\tau \gg \ln k/H].
\end{equation}
The normalisation coefficient in (\ref{eq:normalization}) has been
selected such that
\begin{equation}
    \delta(\omega-\omega') = \int_{-\infty}^{+\infty} d\tau \,
    T_\omega(\tau) T_{\omega'}^*(\tau).
\end{equation}

Putting (\ref{eq:decomposition 1}) into (\ref{eq:EOMs 2}) yields
\begin{subequations}\label{eq:EOMs 3}
\begin{eqnarray}
    \label{eq:EOMs 3a} 0 & = &
    \left[ \di_y^2 - U(y) + \omega^2 H^2
    \right] \phitilde_\omega, \\ \label{eq:EOMs 3b} 0
    & = & [\omega^2 H^2 +\tfrac{9}{4}H^2-\mu^2] \q_\omega -
    \beta (H\ell)^{3/2} \phitilde_{\b\omega}.
\end{eqnarray}
\end{subequations}
The general solution of (\ref{eq:EOMs 3a}) is
\begin{subequations}\label{eq:decomposition 2}
\begin{align}
    \phitilde_\omega(y) & = a^+_\omega \phitilde^+_\omega(y) + a^-_\omega
    \phitilde^-_\omega(y), \\
    \phitilde^+_\omega(y) & = \mathcal{N}^+ (\sinh Hy)^{1/2}
    R_{i\omega-1/2}^\alpha(\cosh Hy), \\ \phitilde^-_\omega(y)
    & = \mathcal{N}^- (\sinh Hy)^{1/2} Q_{i\omega-1/2}^\alpha(\cosh Hy),
\end{align}
\end{subequations}
where
\begin{align}
    \nonumber R_\sigma^\alpha(x) & = \frac{\pi e^{i\pi\alpha} P_\sigma^\alpha(x) -
    Q_\sigma^\alpha(x)\sin\pi(\sigma+\alpha)\sec\pi\sigma
    }{\Gamma(\sigma+3/2)\Gamma(\sigma+1/2)}, \\
    \nonumber
    \mathcal{N}^\pm & = 2e^{-i\pi\alpha}\left( \frac{ \pi\omega  }{H\sinh\pi\omega}
    \right)^{1/2}
    [\Gamma(i\omega+\tfrac{1}{2}\mp\alpha)]^{\pm 1},
    \\ \alpha & = \sqrt{4+m^2\ell^2}.
\end{align}
Here, $P_\sigma^\alpha(x)$ and $Q_\sigma^\alpha(x)$ are associated
Legendre functions \cite[8.7]{GR}.  The coefficients
$\mathcal{N}^\pm$ have been selected such that when we use the
standard asymptotic expansions \cite[8.776]{GR}, we have
\begin{multline}\label{eq:asymptotic behaviour 2}
    T_\omega(\tau)\phitilde^\pm_\omega(y) \sim e^{iH\omega(\tau\pm
    y)}, \\ [\text{$H\tau \gg \ln k/H$ and $Hy \gg 1$}].
\end{multline}
Hence, $\phitilde$ reduces to a superposition of incoming and
outgoing plane waves in the ``asymptotic region'' defined in the
square brackets; i.e., far into the future and far from the brane.

We have yet to enforce the boundary condition (\ref{eq:boundary 1}).
In terms of the mode decomposition, this reads
\begin{equation}\label{eq:boundary 2}
    \left[ \di_y \phitilde_\omega + \tfrac{3}{2} \gamma(H\ell)^{-1} \phitilde_\omega
    \right]_\b = \tfrac{1}{2} \beta(H\ell)^{-3/2} \q_\omega.
\end{equation}
When (\ref{eq:decomposition 2}) is substituted into (\ref{eq:EOMs
3b}) and (\ref{eq:boundary 2}), we obtain a homogeneous system of
two equations linear in $(a^\pm_\omega,\q_\omega)$ which can be
solved for the amplitude ratios $a^\pm_\omega/\q_\omega$.  The
derivatives of Legendre functions appearing in this system can be
simplified using recurrence relations \cite[8.732]{GR}.

We define the scattering matrix (i.e.,~reflection coefficient)
$\mathcal{S}_\omega$ as the ratio of outgoing to ingoing flux in
the asymptotic region (cf.~Eq.~\ref{eq:asymptotic behaviour 2}) at
a particular frequency $\omega$. Explicitly, we obtain
\begin{multline}\label{eq:scattering matrix}
    \mathcal{S}_\omega = \frac{a^-_\omega}{a^+_\omega} = -\frac{1}{\omega \Gamma^2(i\omega)} \frac{\mathcal{N}^+}{\mathcal{N}^-}
    \Bigg[ \frac{\cos \pi(\alpha+i\omega)}{\sinh\pi\omega} \\ +i\pi
    e^{i\pi\alpha}
    \frac{c_+P_{i\omega+\onehalf}^\alpha(\gamma) + c_- P_{i\omega-\onehalf}^\alpha(\gamma)}
    { c_+Q_{i\omega+\onehalf}^\alpha(\gamma) + c_-
    Q_{i\omega-\onehalf}^\alpha(\gamma)} \Bigg],
\end{multline}
where
\begin{subequations}
\begin{align}
    c_+ & = \left(\omega^2+\frac{9}{4}-\frac{\mu^2}{H^2}\right)\left(\alpha-i\omega-\frac{1}{2}\right), \\
    c_- & = \frac{\ell\beta^2}{2H^2}+i\gamma\left(\omega^2+\frac{9}{4}-\frac{\mu^2}{H^2}\right)
    \left(\omega+\frac{3}{2}i\right).
\end{align}
\end{subequations}
Before moving on, we should mention cases for which
(\ref{eq:scattering matrix}) is not a valid expression for the
scattering matrix:
\begin{equation}\label{eq:restrictions}
    i\omega -1/2 \pm \alpha \notin \mathbb{Z}, \quad i\omega \notin
    \mathbb{Z},
\end{equation}
where $\mathbb{Z}$ represents the integers. The first of these
restrictions is in place to ensure that the mode functions
$\phitilde_\omega^\pm$ are linearly independent \cite[8.707]{GR},
while the second is required for the asymptotic expressions
(\ref{eq:asymptotic behaviour 2}) to be applicable \cite[8.776]{GR}.
When $\omega$ is real, both conditions are automatically satisfied.

\subsection{Resonances}

Intuitively, a resonance of a given system is a mode of oscillation
where the response (``output'') is maximised relative to the
stimulus (``input'').  In scattering theory, this implies that the
outgoing flux associated with a resonant mode completely dominates
the ingoing flux.  This leads to the formal definition of a resonant
excitation: A mode whose frequency corresponds to a pole of the
scattering matrix $\mathcal{S}_\omega$ when $\omega$ is continued
into the complex plane \cite{Taylor,LL,Nollert,Clarkson:2005mg}. The
physical importance of these resonant modes is that they describe
the dominant behaviour of any given system, regardless of the choice
of initial conditions. In this context, we expect that the late time
behaviour of the brane and bulk fields to be well approximated by
resonant solutions irrespective of the configuration on some initial
data hypersurface. In this subsection, we study the poles of
$\mathcal{S}_\omega$ in some detail.

\subsubsection{General properties of quasinormal modes}\label{sec:QNM}

\begin{figure}
\begin{center}
\includegraphics[width=\columnwidth]{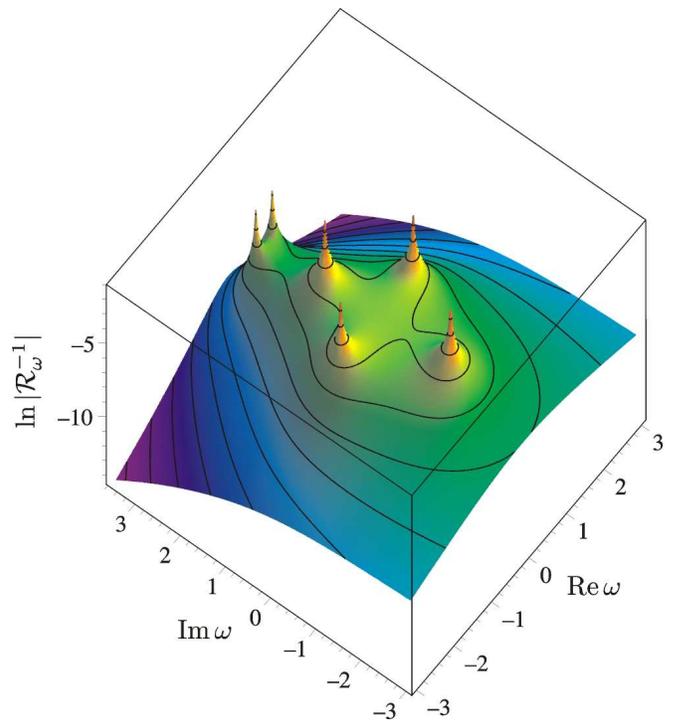}\\
\caption{An example of the distribution of zeros of
$\mathcal{R}_\omega$ in the complex frequency plane. The vertical
axis is $\ln|\mathcal{R}^{-1}_\omega|$, so the tall spikes
represent poles of the scattering matrix.  For this plot, we have
taken $\beta\ell^{3/2} = 1.00$, $m\ell = 3.34$, $H\ell = 1.37$ and
$\mu\ell = 1.00$.}\label{fig:zeros of R}
\end{center}
\end{figure}

When the scattering matrix (\ref{eq:scattering matrix}) is continued
into the complex plane, one discovers that it has a number of poles
on the imaginary $\omega$ axis that violate the restrictions
(\ref{eq:restrictions}).  Since the scattering matrix itself is
undefined at these locations, these poles are ignorable and we omit
them from the subsequent discussion. We can show that all of the
physical poles that remain correspond to roots of
$\mathcal{R}_\omega$, where
\begin{multline}\label{eq:resonance condition}
    \!\!\!\mathcal{R}_\omega = \left(\frac{\mu^2}{H^2} - \omega^2 - \frac{9}{4} \right) \left(
    \alpha - i\omega - \frac{1}{2} \right) Q_{i\omega +
    \tfrac{1}{2}}^\alpha(\gamma)
    \\ + \left[ \gamma \left( i\omega - \frac{3}{2} \right)\left(\frac{\mu^2}{H^2} -
    \omega^2 - \frac{9}{4} \right) - \frac{\ell\beta^2}{2H^2} \right]
    Q_{i\omega - \tfrac{1}{2}}^\alpha(\gamma).
\end{multline}
This condition that $\mathcal{R}_\omega = 0$ is the same as
demanding that a particular mode function satisfies the boundary
condition at the brane and is a purely outgoing wave in the
asymptotic future; i.e., $a^+_\omega = 0$.  Hence, the resonant
states we have defined here are akin to the quasinormal modes of
black hole perturbation theory \cite{Nollert}. In
Fig.~\ref{fig:zeros of R}, we show a plot of
$|\mathcal{R}_\omega^{-1}|$ as a function of complex frequency for
a certain choice of parameters.

It is useful to note that the full spatial and temporal dependance
of a resonant solution for the canonical and physical bulk fields
are
\begin{subequations}
\begin{align}
    \phitildeR_\omega(\tau,y) & = T_\omega(\tau)
    \phitilde_\omega(y), \\
    \phiR_\omega(\tau,y) & = e^{-3H\tau/2} (\sinh Hy)^{3/2} T_\omega(\tau)
    \phitilde_\omega(y),
\end{align}
\end{subequations}
respectively.  In both cases, $\omega$ is a root of
$\mathcal{R}_\omega$. The asymptotic behaviour of these resonant
solutions is crucial to the discussion below.  We have for late
times
\begin{subequations}\label{eq:asymptotic behaviour}
\begin{align}
    \lim_{H\tau\rightarrow\infty} \, & |\phitildeR_\omega(\tau,y)| \propto e^{-\Gamma H \tau},
    \\ \lim_{H\tau\rightarrow\infty} \, & |\phiR_\omega(\tau,y)| \propto e^{-(\Gamma
    +3/2) H\tau}; \\ \intertext{while far from the brane, we obtain}
    \lim_{Hy\rightarrow\infty} \,& |\phitildeR_\omega(\tau,y)|  \propto e^{\Gamma
    Hy}, \\
    \quad \lim_{Hy\rightarrow\infty} \,& |\phiR_\omega(\tau,y)| \propto e^{(\Gamma +
    3/2)Hy},
\end{align}
\end{subequations}
where $\Gamma = \Im\omega$.

We can use this asymptotic behaviour to derive a useful result for
resonances with $\Im\omega < 0$.  We multiply (\ref{eq:EOMs 3a}) by
$\phitilde_\omega^*$ and integrate over $y$. This yields
\begin{equation}\label{eq:proof intermediate}
    -\left[ \phitilde^*_\omega \phitilde'_\omega \right]_\b =
    \int\limits_{y_\b}^{\infty} dy\left\{ |
    \phitilde'_\omega|^2 + [U-\omega^2H^2]|\phitilde_\omega|^2
    \right\},
\end{equation}
where we have integrated by parts and made use of the fact that
$\phitilde_\omega$ vanishes strongly as $y\rightarrow\infty$ if
$\Im\omega < 0$. The lefthand side can be expanded using
(\ref{eq:EOMs 3b}) and the boundary condition (\ref{eq:boundary 2}).
Then, taking the imaginary part of (\ref{eq:proof intermediate}), we
get
\begin{equation}
    0 = \Im(\omega^2) \left[ |\q_\omega|^2 + 2(H\ell)^{3}
    \int_{y_\b}^{\infty} dy |\phitilde_\omega|^2
    \right].
\end{equation}
Since the quantity inside the square brackets is manifestly nonzero,
we are led to the conclusion $\Im(\omega^2)=0$.  But, we also have
that $\Im\omega \ne 0$, from which it follows that $\Re\omega = 0$.
That is, any resonant frequencies $\omega$ satisfying
$\mathcal{R}_\omega = 0$ in the lower half of the complex plane must
lie on the imaginary axis.

\subsubsection{Classification of resonances}

We now turn to the classification of resonant modes with complex
frequency. One important way of discriminating between different
classes of resonance is the asymptotic behaviour articulated in
Eq.~(\ref{eq:asymptotic behaviour}).  Another categorization scheme
involves assigning a ``Kaluza-Klein'' mass $\rho$ to each resonance,
which is implicitly defined by noting that $\phiR_\omega$ satisfies
the Klein-Gordon equation on the brane
\begin{equation}
    -\rho^2\phiR_\omega = (\di_\tau^2 + 3H \, \di_\tau + k^2
    e^{-2H\tau})\phiR_\omega,
\end{equation}
where the field mass is
\begin{equation}\label{eq:KK mass}
    \rho^2 = H^2\left( \omega^2 + \tfrac{9}{4}
    \right).
\end{equation}
Note that for $\Im\omega < 0$, $\rho$ will be purely real or
imaginary.  On the other hand, if $\Im\omega > 0$ the mass will be
complex in general.  Finally, we can also classify resonances
based on whether or not they are normalisable under the
Klein-Gordon inner product:
\begin{equation}\label{eq:KG inner product}
    (\phi,\psi)_\KG = \frac{1}{V_3} \int_{\Sigma_\tau} t^a(\phi^*
    \di_a \psi - \psi \di_a \phi^*),
\end{equation}
where $\Sigma_\tau$ is the spacelike 4-surface of constant $\tau$
and $t^a$ is its future directed normal vector.  We apply a
box-normalization of volume $V_3$ for the spatial 3-manifold.
Consider the norm of $\phiR_\omega$
\begin{eqnarray}
    \!\! (\phiR_\omega,\phiR_\omega)_\KG & \propto &
    \int\limits_{y_\b}^\infty dy \frac{|\phiR_\omega|^2}{(\sinh
    Hy)^3} \propto \int\limits_{y_\b}^\infty dy \, |\phitildeR_\omega|^2.
\end{eqnarray}
Hence, $\phiR_\omega$ is normalisable if $\phitildeR_\omega$ is
square integrable over $y$.

\begin{figure}
\begin{center}
\includegraphics{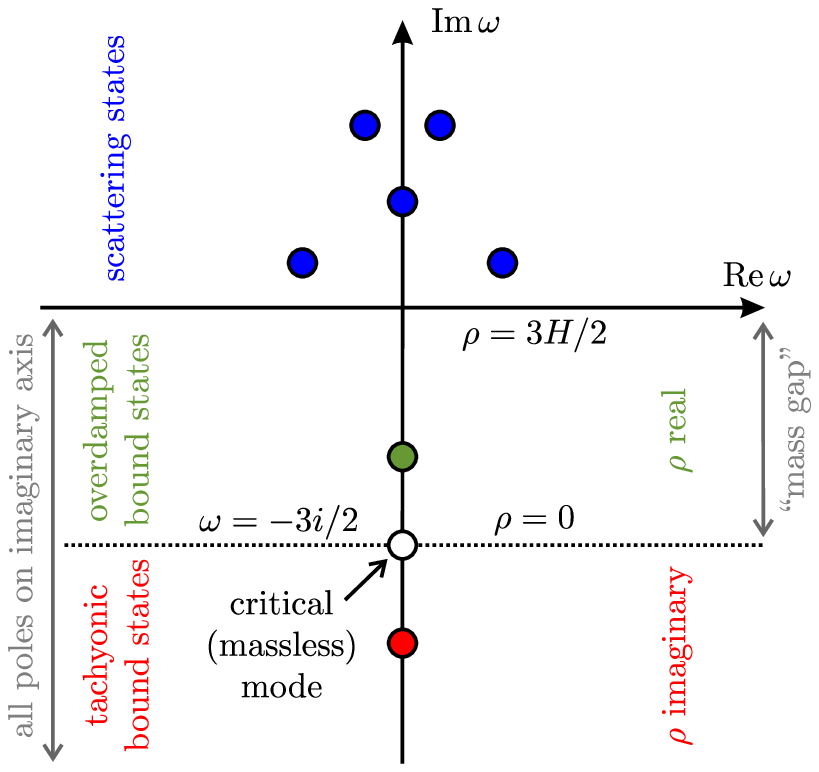}\\
\caption{Different classes of possible resonances. The circles
indicate poles of the scattering matrix.}\label{fig:classification}
\end{center}
\end{figure}

Armed with these definitions, we can delineate states into three
categories:

\begin{description}

\item[Scattering states:\,] These are resonances with $\Im\omega > 0$.
For these modes, both $\phitildeR_\omega$ and $\phiR_\omega$ vanish
at late times. Conversely, both states also diverge for large $y$,
and the Klein-Gordon norm $(\phiR_\omega,\phiR_\omega)_\KG$
diverges. Hence, these are \emph{not} conventional bound states.
This spatial divergence does not mean that scattering states are
unphysical, since they are well behaved if the limits $\tau
\rightarrow \infty$ and $y \rightarrow \infty$ are taken
simultaneously. Scattering states are akin to quasinormal modes in
black hole perturbation theory in that they are never observed at
spatial infinity, only future null infinity.  The Kaluza-Klein mass
of scattering states is generally complex.

\item[Overdamped bound states:\,]  These have $-3/2 < \Im\omega < 0$
which implies $\Re\omega=0$.  At late times, the canonical field
$\phitildeR_\omega$ blows-up while the physical field
$\phiR_\omega$ is exponentially damped.  These modes are
normalisable under the Klein-Gordon product since
$(\phiR_\omega,\phiR_\omega)_\KG$ is finite.  The Kaluza-Klein
mass of these modes is real since $\rho^2
> 0$.  Because $\phiR_\omega$ vanishes in the infinite future and is
normalisable, we call these modes ``overdamped bound states''.

\item[Tachyonic bound states:\,]  These have $\Im\omega < -3/2$ and
$\Re\omega = 0$.  They are normalisable and diverge at late times.
In this case, the Kaluza-Klein mass is imaginary $\rho^2 < 0$, which
is why we call these modes ``tachyonic bound states''.

\end{description}
We summarize this classification scheme in
Fig.~\ref{fig:classification}.  We have indicated the ``mass gap''
between $\rho = 0$ and $\rho = 3H/2$ familiar from the analysis of
uncoupled bulk fields about de Sitter branes.

\subsubsection{Critical coupling and the massless mode}\label{sec:critical coupling}

For any choice of $\mu$, $m$, $H\ell>1$, it is always possible to
choose the coupling parameter $\beta = \betacrit \in \mathbb{R}$
such that there is one resonance with frequency $\omega = -3i/2$
and Kaluza-Klein mass $\rho=0$. To see this, we substitute $\omega
= -3i/2$ in $\mathcal{R}_\omega = 0$ and solve for $\betacrit$:
\begin{equation}\label{eq:critical coupling}
    \betacrit = \frac{\mu \sqrt{2(\alpha-2)}}{\ell^{1/2}} \left[
    \frac{Q_2^\alpha(\gamma)}{Q_1^{\alpha}(\gamma)}\right]^{1/2}.
\end{equation}
Note that a fundamental property of the associated Legendre
function of the second kind is that \cite[8.783]{GR}
\begin{equation}\label{eq:Q zeroes}
    e^{-i\pi\alpha} Q_\sigma^\alpha(\gamma) > 0 \,\text{ for all }\, \sigma >
    -\tfrac{3}{2},\,\,
    \alpha > -\tfrac{1}{2},\,\, \gamma > 1.
\end{equation}
From this it follows that $\betacrit$ is real for all possible
parameter choices; i.e., it is always possible to select the
coupling such that at least one resonance is massless. Note also
that $\betacrit = 0$ when either $m$ or $\mu$ is zero.

The qualitative behaviour of this massless mode at late times is
the same a massless scalar propagating in 4-dimensional de Sitter:
Namely, the physical field amplitude $\phiR_\omega$ is frozen
(conserved) on superhorizon scales.

\subsubsection{Bound state resonances}\label{sec:bound states}

\begin{figure*}
\begin{center}
\subfigure[\,\,A few of the complex roots of $\mathcal{R}_\omega$ as
functions of the coupling parameter
$\beta$]{\includegraphics[scale=1]{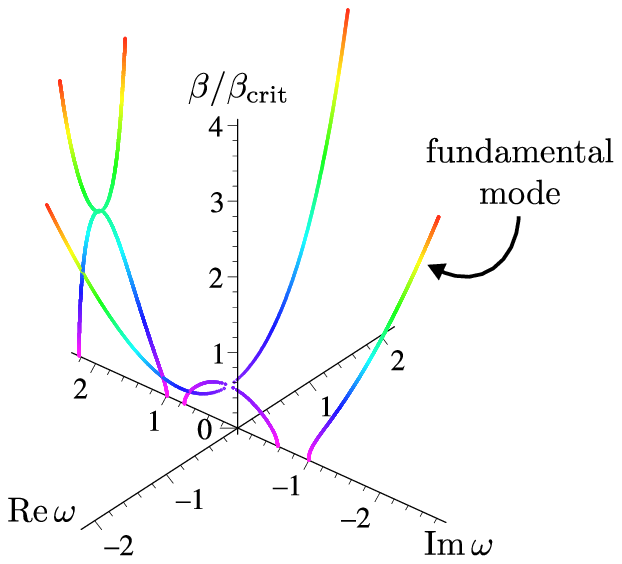}}\hspace{0.3in}
\subfigure[\,\,Variation with $\beta/\betacrit$ in the imaginary
parts of the resonant frequencies shown in
(a)]{\includegraphics{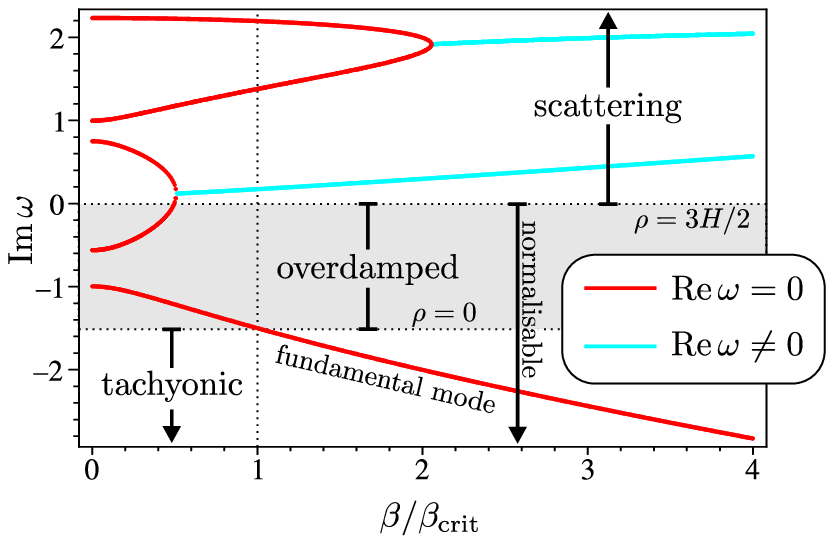}} \caption{The least-damped
resonant frequencies of the system when $\mu\ell = 1.40$, $m\ell =
2.37$ and $H\ell = 1.25$.  This particular example illustrates the
generic feature of our model: a tachyonic resonance only exists for
supercritical ($\beta
> \betacrit$) coupling. Note that all
resonances with $\Im\omega <0$ have $\Re\omega=0$, as expected. Also
note that the two $\beta = 0$ resonances $\omega =\pm 0.998i$ are
modes for which the bulk field vanishes $\phiR_\omega = 0$
(i.e.~these are pure $q$ modes), while the other modes have a
vanishing brane mode $q_\omega^\text{\tiny{R}}=0$.  The
``fundamental mode'' has been identified as the resonance with the
smallest imaginary part, which is expected to dominate the late time
behaviour of numeric simulations
(cf.~Sec.~\ref{sec:results}).}\label{fig:more zeros of R}
\end{center}
\end{figure*}

\begin{figure}
\begin{center}
\includegraphics[scale=0.95]{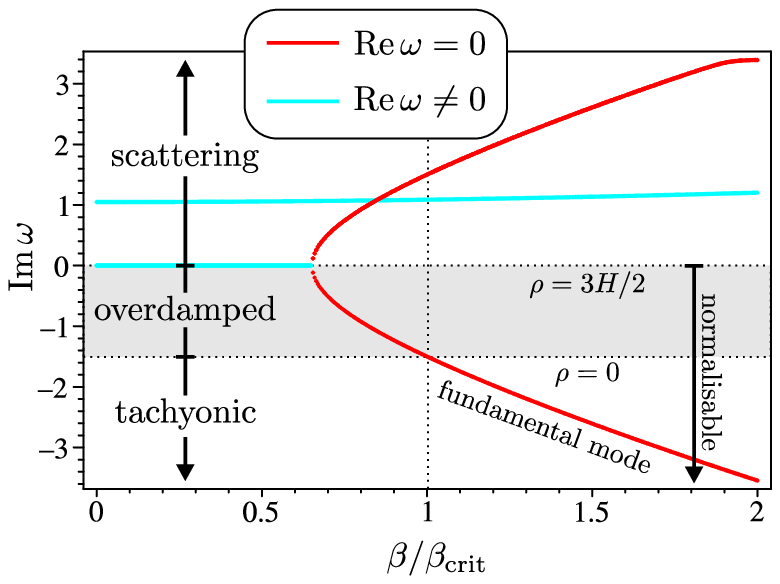}\\
\caption{Variation in the imaginary parts of the least-damped
resonant frequencies with coupling when $\mu\ell = 0.40$, $m\ell =
2.1$ and $H\ell=0.20$.  In contrast to Fig.~\ref{fig:more zeros of
R}, there is no bound state at small coupling.  For
$\beta/\betacrit \gtrsim 0.65$, there is a single bound state that
is overdamped or tachyonic for subcritical or supercritical
coupling, respectively.}\label{fig:even more zeros of R}
\end{center}
\end{figure}

In Fig.~\ref{fig:more zeros of R}, we track the resonant roots of
$\mathcal{R}_\omega$ through the complex plane as a function of the
coupling $\beta/\betacrit$ for particular values of the masses. Here
and in Sec.~\ref{sec:results} below, the roots are found numerically
using Newton's method.  We see that there are bound states for all
couplings, but there is only a tachyonic mode for supercritical
($\beta > \betacrit$) coupling. Furthermore, in the small coupling
regime, we see two overdamped bound states. The behaviour of the
resonant frequencies for a different choice of parameters is shown
in Fig.~\ref{fig:even more zeros of R}.  In this scenario, no bound
state even exists for small coupling.  In both cases, we see that
the fundamental mode goes from overdamped to tachyonic as the
coupling is increased through the critical value.  There is no mode
mixing in general at this coupling, although at other couplings it
is possible for modes to merge or bifurcate (for example, around
$\beta \sim 0.6 \times \betacrit$ in Fig.~\ref{fig:more zeros of
R}). Our purpose in this subsection is to understand some of this
bound state behaviour in terms of the general properties of
$\mathcal{R}_\omega$.

\paragraph*{Massive bulk and brane fields}

We are only interested in the bound states, so we set
\begin{equation}\label{eq:lambda def}
    \omega = -i(\lambda+\tfrac{3}{2}), \quad \lambda >
    -\tfrac{3}{2}.
\end{equation}
If the Kaluza-Klein mass of a resonance corresponding to a given
value of $\lambda$ is $\rho_\lambda$, then we have
\begin{equation}
    \rho_\lambda^2 = -H^2\lambda(\lambda+3) =
    \begin{cases}
        \text{positive}, & \lambda \in [-\tfrac{3}{2},0), \\
        \text{negative}, & \lambda \in (0,\infty);
    \end{cases}
\end{equation}
i.e., any bound state with $\lambda >0$ is a tachyon.

Assuming that $\alpha > 2$ and $\mu > 0$, and mindful of (\ref{eq:Q
zeroes}), we can re-write the resonance condition as
\begin{equation}
    (\beta/\betacrit)^2 =
    \mathcal{Z}(\lambda),
\end{equation}
where
\begin{multline}
    \mathcal{Z}(\lambda) =
    \frac{1}{(\alpha-2)} \frac{Q_1^\alpha(\gamma)}
    {Q_2^\alpha(\gamma)} \left(
    1 - \frac{\rho_\lambda^2}{\mu^2}
    \right) \times \\ \left[ (\lambda + \alpha + 1)
    \frac{Q_{\lambda}^\alpha(\gamma)}
    {Q_{\lambda+1}^\alpha(\gamma)} - \gamma(\lambda+3)
    \right].
\end{multline}
Empirically, we find that for all $\alpha > 2$, $\mu > 0$ and
$\gamma > 1$, the continuously differentiable function
$\mathcal{Z}(\lambda) \in \mathbb{R}$ has the following
properties:
\begin{itemize}
    \item Over the interval $\lambda \in [0,\infty)$,
    $\mathcal{Z}(\lambda)$ monotonically increases from $1$ to
    $\infty$.  Therefore, there always exists one and only one
    tachyonic bound state for supercritical coupling $\beta > \betacrit$.
    \item In the interval $\lambda \in [-3/2,0]$,
    $\mathcal{Z}(\lambda)$ has at most one local minimum and no
    local maxima.  Therefore, for any choice of parameters we can
    have at most two bound state resonances.
    \item The absolute minimum of $\mathcal{Z}(\lambda)$ for
    $\lambda \in [-3/2,\infty)$ is always less than unity, but not necessarily
    less than zero.  Therefore, it is possible to select parameters such
    that there are no bound states at all.  Such situations will
    always have subcritical coupling $\beta < \betacrit$.
\end{itemize}
We reiterate that these conclusions are only valid with both
fields have finite mass.

From these comments, we can deduce that a necessary (but not
sufficient) condition for the existence of two bound states is
\begin{equation}
    \mathcal{Z}(-3/2) > 0 \text{ and } \mathcal{Z}'(-3/2) < 0.
\end{equation}
By direct sampling of the $(\mu/H,m\ell,H\ell)$ parameter space, we
have numerically determined that these conditions hold if and only
if
\begin{equation}\label{eq:bound state inequalities}
    \mu/H < 3/2 \text{ and } m^2/H^2 <
    \hat{m}_\text{thres}^2(H\ell).
\end{equation}
Here, the dimensionless threshold mass
$\hat{m}_\text{thres}^2(H\ell)$ is obtained numerically and shown
in Fig.~\ref{fig:threshold mass}.  We will see in
Sec.~\ref{sec:zero coupling} below, that the second inequality in
(\ref{eq:bound state inequalities}) is a necessary and sufficient
condition for there to exist a bound state of $\phi$ at zero
coupling.
\begin{figure}
\begin{center}
\includegraphics[width=\columnwidth]{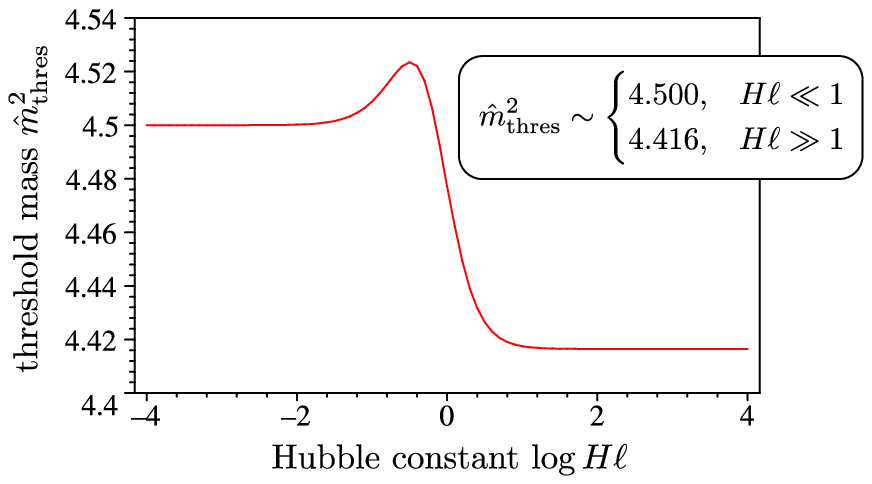}\\
\caption{The threshold mass as obtained from numeric calculations.
The condition $m^2/H^2 < \hat{m}^2_\text{thres}(H\ell)$ is
necessary for the existence of two bounds states when the coupling
is nonzero. In Sec.~\ref{sec:zero coupling}, we also see that this
condition is necessary and sufficient for the existence of a bound
state of the bulk field at zero coupling.}\label{fig:threshold
mass}
\end{center}
\end{figure}

\paragraph*{Massless bulk field}

The situation is somewhat different when either $m$ or $\mu$ is
set to zero.  For the sake of brevity, we only mention the case of
the massless bulk field here.  When $\alpha = 2$, the condition
for bound states is
\begin{equation}
    \beta^2/H^3 = \mathcal{Y}(\lambda),
\end{equation}
where,
\begin{equation}
    \mathcal{Y}(\lambda) = \frac{2\rho_\lambda^2(\rho_\lambda^2-\mu^2)}{H^4}
    \left[ -\frac{Q_{\lambda+1}^1(\gamma)}
    {Q_{\lambda+1}^2(\gamma)} \right].
\end{equation}
The function $\mathcal{Y}(\lambda) \in \mathbb{R}$ is continuously
differentiable for $\lambda > -3/2$, and has the following
properties:
\begin{itemize}
    \item $\mathcal{Y}(\lambda)$ monotonically increases from $0$ to
    $\infty$ for $\lambda \in [0,\infty)$.  Therefore, there always
    exists one and only one tachyonic bound state for any nonzero
    coupling $\beta > 0$.
    \item The sign of $\mathcal{Y}$ at $\lambda = -3/2$ satisfies
    $\text{sgn}\,[\mathcal{Y}(-3/2)] =
    \text{sgn}\,[\tfrac{3}{2}H-\mu]$.  Therefore, in addition to
    the ubiquitous tachyonic mode, one can have an
    overdamped bound state if and only if the mass of the brane
    field lies in the mass gap $\mu < \tfrac{3}{2}H$.
    \item $\mathcal{Y}(\lambda)$ has at most one local extremum for
    $\lambda \ge -3/2$.  If this extremum exists, it must be a minimum
    inside $[-3/2,0]$. Hence, we can either have one tachyonic and one overdamped
    resonance, or one tachyonic resonance only.
\end{itemize}

\subsection{Limiting cases}\label{sec:limiting cases}

In this subsection, we consider the resonance condition
$\mathcal{R}_\omega = 0$ in various limits and hence recover several
results from the literature.

\subsubsection{Zero coupling}\label{sec:zero coupling}

We first consider the zero coupling limit of the model $\beta = 0$.
It is useful to define
\begin{subequations}
\begin{align}\nonumber
    \mathcal{R}_\omega^{(\phi)} & = \left(
    \alpha - i\omega - 1/2 \right) Q_{i\omega +
    1/2}^\alpha(\gamma)
    + \\  \label{eq:zero coupling a} & \qquad \qquad \qquad \qquad
    \gamma \left( i\omega - 3/2 \right)
    Q_{i\omega - 1/2}^\alpha(\gamma), \\
    \label{eq:zero coupling b} \mathcal{R}_\omega^{(q)} & =
    {\mu^2}/{H^2} - \omega^2 - 9/4.
\end{align}
\end{subequations}
In terms of these functions, we find that $\mathcal{R}_\omega = 0$
implies
\begin{equation}
    \mathcal{R}_\omega^{(\phi)} = 0 \text{ or }
    \mathcal{R}_\omega^{(q)} = 0, \quad [\beta = 0].
\end{equation}
Going back to the analysis of Sec.~\ref{sec:scattering matrix}, we
can show that if either the first or second condition holds, then
$\tilde{q}_\omega^\text{\tiny R}(\tau)=0$ or
$\phitildeR_\omega(\tau,y) = 0$, respectively.  In other words, each
condition represents the resonances of one field decoupled from the
other.  The brane field resonances are trivial
\begin{equation}
    \omega = \pm\sqrt{\mu^2/H^2 - 9/4} \quad \Rightarrow \quad \rho^2 =
    \mu^2;
\end{equation}
that is, these modes oscillate with the natural frequency of the
brane oscillator in isolation and have a Kaluza-Klein mass equal
to $\mu$.

Using an identity \cite[8.739]{GR} and recurrence relations
\cite[8.73]{GR}, the bulk resonance condition
$\mathcal{R}_\omega^{(\phi)}=0$ can be re-written as
\begin{align}
    \nonumber 0 & = \zeta (\alpha-2) P_{\alpha-1/2}^{-i\omega}(\zeta) +
    (i\omega-\alpha+\tfrac{1}{2})
    P_{\alpha-3/2}^{-i\omega}(\zeta) \\ & = (i\omega+\alpha+\tfrac{1}{2})
    P_{\alpha+1/2}^{-i\omega}(\zeta) - \zeta(\alpha+2)P_{\alpha-1/2}^{-i\omega} (\zeta),
\end{align}
where $\zeta = \gamma/(H\ell)$.  These match the various bound
state criteria obtained for a massive scalar field about a de
Sitter brane in
Refs.~\cite{Sago:2001gi,Himemoto:2000nd,Langlois:2003dd} (up to
trivial differences in notation).

As in Sec.~\ref{sec:bound states} above, we can substitute in
$\omega = -i(\lambda+\tfrac{3}{2})$ into the bulk resonance
condition $\mathcal{R}_\omega^{(\phi)} = 0$ to derive a criterion
for bound state resonances of $\phi$ at zero coupling. This is
\begin{equation}
    0 = \mathcal{X}(\lambda) =
    \frac{Q_{\lambda+2}^\alpha(\gamma)}{Q_{\lambda+1}^\alpha(\gamma)}
    - \frac{\gamma\lambda}{\lambda+2-\alpha}, \quad \lambda >
    -\tfrac{3}{2}.
\end{equation}
One can confirm that $\mathcal{X}(\lambda)\in\mathbb{R}$ is
monotonically increasing for $\lambda > -3/2$ and positive
definite for $\lambda > 0$.  Therefore, a de-coupled massive bulk
field can exhibit at most one bound state, and this must be an
overdamped resonance.  Furthermore, a necessary and sufficient
condition for the existence of such a bound state is that
$\mathcal{X}(-3/2) < 0$, which we have numerically determined is
equivalent to
\begin{equation}
    m^2/H^2 < \hat{m}^2_\text{thres}(H\ell),
\end{equation}
where the function $\hat{m}^2_\text{thres}(H\ell)$ is shown in
Fig.~\ref{fig:threshold mass}.  In other words, if the bulk field
is too heavy compared with the Hubble scale, it cannot be
effectively localised near the brane.

Finally, setting the bulk field mass to zero (i.e.~$\alpha = 2$) in
(\ref{eq:zero coupling a}) yields that
\begin{equation}
    0 = \left( \omega+\tfrac{3}{2}i \right) \left( \omega-\tfrac{3}{2}i \right)
    Q_{i\omega - 1/2}^1(\gamma).
\end{equation}
In this case there is always a critical bound state with $\omega =
-3i/2$, which is usually called the zero mode; i.e.,
\begin{equation}\label{eq:zero mode}
    \text{zero mode} \,\, \Leftrightarrow \,\, \omega = -3i/2, \,\,
    m = 0, \,\, \beta = 0.
\end{equation}
One can confirm that for the zero mode, the resonant solution is
$y$-independent
\begin{equation}\label{eq:zero mode profile}
    \di_y \phiR(\tau,y) = 0.
\end{equation}
This fact will be useful in the interpretation of the effective
4-dimensional theory in Sec.~\ref{sec:effective theory}.

\subsubsection{Perturbed zero mode}\label{sec:perturbed zero mode}

Since the coefficients in front the Legendre functions in
(\ref{eq:resonance condition}) vanish when $m = 0$ and $\omega =
-3i/2$, it is possible to linearise the resonance condition about
the zero mode solution (\ref{eq:zero mode}).  In particular, we
can take
\begin{equation}\label{eq:zero mode limits}
    |\omega+\tfrac{3}{2}i| \ll 1, \quad m\ell \ll 1, \quad \ell\beta^2 \ll \min(\mu^2,H^2),
\end{equation}
expand $\mathcal{R}_\omega$ to linear order in small quantities, and
solve for $\omega$.  Substituting this solution into (\ref{eq:KK
mass}), we obtain
\begin{align}\label{eq:zero mode mass}\nonumber
    \delta\rho^2_\text{zero} & \sim  m^2 g(H\ell) -
    \ell^{-1}\mu^{-2}\beta^2 f(H\ell)
    \\ & \sim m^2 g(H\ell) ( 1 -
    {\beta^2}/{\betacrit^2} ),
\end{align}
where $\delta\rho_\text{zero}$ is the small Kaluza-Klein mass that
the zero mode acquires when either $m$ or $\beta$ is nonzero. Here,
we have defined
\begin{subequations}\label{eq:f and g def}
\begin{align}
    f(x) & = \left[ \sqrt{x^2+1} - x^2 \ln \left( \frac{1+\sqrt{x^2+1}}{x}
    \right)\right]^{-1}, \\ g(x) & = \frac{1}{2} f(x)\sqrt{x^2+1} -
    \frac{3}{4} x^2,
\end{align}
\end{subequations}
and made use of the limiting values of the critical coupling
(\ref{eq:critical coupling}) for $m\ell \ll 1$:
\begin{equation}\label{eq:critical coupling approx}
    \betacrit^2 \sim \ell m^2 \mu^2 \frac{g(H\ell)}{f(H\ell)} \rightarrow
    \begin{cases}
        \tfrac{1}{2} m^2\mu^2\ell, & H\ell \ll 1, \\
        \tfrac{2}{5} m^2\mu^2H^{-1}, & H\ell \gg 1,
    \end{cases}
\end{equation}
as well as explicit representations of the Legendre functions with
integral indices \cite[8.827]{GR}. We see that the mass of the bulk
field tends to increase the squared mass of the zero mode into the
overdamped region, while the coupling tends to decrease it into the
tachyonic regime.  We also see that supercritical coupling
$\beta>\betacrit$ explicitly leads to a tachyonic mode
$\delta\rho_\text{zero}^2 < 0$.  For reference, we note the limiting
behaviour of $\delta\rho_\text{zero}$ when $H\ell$ takes on extreme
values:
\begin{equation}
    \delta\rho^2_\text{zero} \rightarrow
    \begin{cases}
        \tfrac{1}{2}m^2 - \ell^{-1}\mu^{-2}\beta^2, & H\ell \ll 1, \\
        \tfrac{3}{5}m^2 - \tfrac{3}{2}\mu^{-2}H\beta^2, & H\ell \gg 1.
    \end{cases}
\end{equation}
Note that for the perturbed zero mode to be a bound state, we need
\begin{equation}
    \delta\rho_\text{zero} < 3H/2 \,\, \Rightarrow \,\, m/H <
    \sqrt{3}/2, \quad [H\ell \ll 1].
\end{equation}
This matches the small $H\ell$ behaviour of the bound state
condition $m^2/H^2 < \hat{m}_\text{thres}^2$, where
$\hat{m}_\text{thres}^2$ is shown in Fig.~\ref{fig:threshold
mass}.  There is no conclusion to be drawn in the other extreme,
because the conditions (\ref{eq:zero mode limits}) imply
$\delta\rho_\text{zero} / H \ll 1$ for $H\ell \gg 1$; i.e., the
perturbed zero mode is always a bound state.

We note that the expression (\ref{eq:zero mode mass}) for
$\delta\rho_\text{zero}$ closely resembles that obtained by
\citet{Langlois:2003dd}, who studied a massive bulk scalar with a
brane-localized self-coupling. Their model was governed by the
action
\begin{equation}
    S_\text{\tiny LS} = -\int\limits_\mathcal{M} \left( g^{ab}
    \, \di_a \phi \, \di_b \phi + m^2
    \phi^2 \right) -
    \int\limits_{\di\mathcal{M}_\text{b}} \! \frac{\chi}{\ell}
    \phi^2,
\end{equation}
where $\chi$ is a dimensionless constant parameter. Their results
for the perturbative zero mode mass match ours if we make the
identification
\begin{equation}
    \chi = -\tfrac{1}{2}\mu^{-2}\beta^2\ell.
\end{equation}
Hence, under the conditions (\ref{eq:zero mode limits}), our system
behaves like a massive bulk scalar field with a constant brane
self-coupling.

\subsubsection{Low energy}\label{sec:low energy}

The low energy limit of the model is defined by
\begin{equation}
    H\ell \ll 1, \quad \gamma \sim 1, \quad \text{[$\ell$ finite]}.
\end{equation}
Consider the temporal eigenvalue equation (\ref{eq:temporal
eigenvalue}) in this regime:
\begin{equation}\label{eq:low energy eigenvalue}
    (\di_\tau^2 + k^2) T_\omega(\tau) = -\omegaphys^2
    T_\omega(\tau), \quad \omegaphys = \omega H,
\end{equation}
where we have identified the dimensionful physical frequency
$\omegaphys$.  When we take the limit $H\ell \rightarrow 0$, we
want to ensure that the physical frequency is finite.  Hence, we
define $\rhohat^2/\ell^2$ to be the \emph{finite} eigenvalue of
$(\di_\tau^2 + k^2)$:
\begin{equation}\label{eq:rhohat def}
    \rhohat = \omega H\ell = \ell \omegaphys, \,\,\, \lim_{H\ell \rightarrow 0}
    |\rhohat| < \infty, \,\,\, \lim_{H\ell \rightarrow 0}
    |\omega| = \infty.
\end{equation}
In terms of $\rhohat$, the solution of (\ref{eq:low energy
eigenvalue}) is
\begin{equation}
    T_\omega(\tau) = e^{i\varpi \tau/\ell}, \quad \varpi = \sqrt{\rhohat^2 + k^2
    \ell^2}.
\end{equation}
To find the resonant modes in this case, our strategy will to be
to take the limit of $\mathcal{R}_\omega = 0$ when $\gamma
\rightarrow 1$, $H\ell \ll 1$, $|\omega| \gg 1$, and $\omega$
real. To avoid branch cut ambiguities, we put $\rhohat \rightarrow
\rhohat - i0^+$ with $\rhohat \in \mathbb{R}$. The resulting
expression can then be continued into the $\rhohat$ complex plane.

Consider the following integral representation of the associated
Legendre function \cite[8.711(4)]{GR}:
\begin{multline}\label{eq:integral representation}
    Q^\alpha_\sigma(\gamma) = \frac{e^{i\pi\alpha}\Gamma(\sigma+1)}
    {\Gamma(\sigma-\alpha+1)} \int_0^\infty \frac{dt\,\cosh \alpha t}
    {(\gamma + \sqrt{\gamma^2-1}\cosh t)^{\sigma+1}}, \\
    [\text{$\gamma > 1$, $\Re(\sigma-\alpha) > -1$, and $-\sigma \ne
    1,2,\ldots$}].
\end{multline}
Now, if we assume $\gamma \rightarrow 1$ and $\sigma = i\omega \pm
1/2 \sim i(\rhohat-i0^+)(H\ell)^{-1}$ in the integrand of
(\ref{eq:integral representation}), we obtain
\begin{multline}\label{eq:integral representation 2}
    Q^\nu_{i\omega + 1/2}(\gamma) \sim \frac{e^{i\pi\nu}\Gamma(i\omega + 3/2)}
    {\Gamma(i\omega -\nu+3/2)} \\ \times \underbrace{\int_0^\infty dt\,
    \exp\left[-i(\rhohat-i0^+)\cosh t \right] \cosh\nu
    t}_{= -\frac{1}{2}i\pi e^{-i\nu\pi/2} \H_\nu^{(2)}(\rhohat -
    i0^+)} + \mathcal{O}(H\ell),
\end{multline}
where the integral \cite[3.457(4)]{GR} has been evaluated assuming
the principal branch of the Hankel functions $\H_\nu^{(1,2)}(z)$
\cite[8.476(7)]{GR} and fundamental definitions \cite[8.407]{GR}.
Using (\ref{eq:rhohat def})--(\ref{eq:integral representation 2}),
a recurrence relation \cite[8.734(4)]{GR}, and expanding to
leading order in $H\ell$, we obtain the resonant condition for low
energies:
\begin{multline}\label{eq:low energy resonance}
    0 = \left[ \tfrac{1}{2} \beta^2\ell^3 + (\alpha-2)(\rhohat^2 -
    \mu^2\ell^2) \right] \H^{(2)}_\alpha(\rhohat-i0^+) \\ -
    \rhohat(\rhohat^2-\mu^2\ell^2) \H^{(2)}_{\alpha-1}(\rhohat-i0^+)
    + \mathcal{O}(H\ell).
\end{multline}
This matches the result obtained in \citet{Koyama:2005gh} if one
takes into account different conventions for the field time
dependence and an implicit choice of integration contour for the
inverse Fourier transform. Note that in this limit, the
Kaluza-Klein mass (\ref{eq:KK mass}) of a given resonance is
\begin{equation}
    \rho^2 = \frac{\rhohat^2}{\ell^2} + \frac{9H^2}{4}
    \xrightarrow[0]{\,\,\,H\ell\,\,\,} \frac{\rhohat^2}{\ell^2}.
\end{equation}
Hence, the critical massless mode corresponds to $\rhohat = 0$.  As
before, this mode is a resonance if $\beta$ equals some critical
value
\begin{equation}\label{eq:low energy critical coupling}
    \betacrit = \mu\ell^{-1/2} \sqrt{2(\alpha-2)}.
\end{equation}
As discussed in detail in Ref.~\cite{Koyama:2005gh}, bound states
in this limit correspond to resonances with $\rhohat$ on the
negative imaginary axis.  One can only have a bound state for
supercritical coupling, and that unique bound state is necessarily
tachyonic.

Eq.~(\ref{eq:low energy resonance}) can be linearised about the
resonance (\ref{eq:zero mode}), resulting in the mass perturbation
of the zero mode:
\begin{equation}
    \delta\rho_\text{zero}^2 \sim \frac{1}{2} m^2 - \frac{1}{\mu^2\ell}
    \beta^2 \sim \frac{1}{2}m^2
    \left(1-\frac{\beta^2}{\betacrit^2}\right),
\end{equation}
where the last approximation comes from expanding the critical
coupling about $m=0$.  One can confirm that this reproduces the
$H\ell \ll 1$ limit of Eq.~(\ref{eq:zero mode mass}).

We can also take the limit of (\ref{eq:low energy resonance}) in the
case of a flat bulk
\begin{equation}
    \ell \rightarrow \infty, \quad \rhohat = \rho\ell \rightarrow \infty,
    \quad m\ell \sim \alpha \rightarrow \infty,
\end{equation}
by making use of the following fact \cite{Koyama:2005gh}:
\begin{equation}
    \frac{\rhohat
    \text{H}^{(2)}_{\alpha-1}(\rhohat)}{\text{H}^{(2)}_{\alpha}(\rhohat)}
    + 2 - \alpha \xrightarrow[\infty]{\,\,\,\,(\rhohat,\alpha)\,\,\,\,}
    -\sqrt{m^2\ell^2-\rhohat^2},
\end{equation}
We obtain
\begin{equation}\label{eq:flat space resonance}
    0 = \beta^2 + 2\sqrt{m^2-\rho^2} (\rho^2 -
    \mu^2), \quad \rho = \rhohat/\ell,
\end{equation}
which matches the resonance condition found by \citet{George:2004wh}
for the case of a Minkowski brane in a Minkowski bulk (with an
appropriate change of notation $\rho^2 \rightarrow \rho^2+m^2$ and
choice of branch $\sqrt{-\rho^2}=-i\rho$).

\subsubsection{High energy}

The high-energy limit is
\begin{equation}\label{eq:high energy limit}
    H\ell \gg 1, \quad \gamma \sim H\ell \gg 1.
\end{equation}
Unlike the low energy limit above, we make no assumptions about
the magnitude of $\omega$.  Using the asymptotic expansion for
$Q_{i\omega\pm 1/2}^\alpha(\gamma)$ for large $\gamma$
\cite[8.776(2)]{GR}, and working to lowest order in
$(H\ell)^{-1}$, we find that the resonant condition
$\mathcal{R}_\omega = 0$ reduces to:
\begin{multline}\label{eq:high energy polynomial}
    \!\!\!\!\! \frac{\beta^2}{H^3} = \frac{1}{i\omega+1} \left(
    \omega^2+\frac{9}{4}- \frac{\mu^2}{H^{2}}\right)
    \left(2\omega^2 + i\omega + 3- \frac{m^2}{H^{2}}
    \right), \\ [\text{$\mu/H$ and $m\ell$ non-infinite}].
\end{multline}
Note that the resonance condition is independent of $\ell$, which is
expected because the characteristic curvature scale on the brane
$H^{-1}$ is much smaller than the characteristic curvature scale of
the bulk $\ell$; i.e., the brane ``sees'' the bulk as flat.

Eq.~(\ref{eq:high energy polynomial}) implies that there are four
resonant frequencies that are the solutions of a quartic polynomial.
The critical coupling is again obtained by setting $\omega = -3i/2$:
\begin{equation}
    \betacrit^2 = \tfrac{2}{5}m^2\mu^2 H^{-1},
\end{equation}
which matches the $H\ell \rightarrow \infty$ limit of
Eq.~(\ref{eq:critical coupling}).  As in Sec.~\ref{sec:perturbed
zero mode}, we can linearise the resonance condition about the zero
mode (\ref{eq:zero mode}), resulting in the perturbation of its
Kaluza-Klein mass:
\begin{equation}
    \delta\rho_\text{zero}^2 \sim \frac{3}{5} m^2 - \frac{3H}{2\mu^2}
    \beta^2 = \frac{3}{5}m^2
    \left(1-\frac{\beta^2}{\betacrit^2}\right).
\end{equation}
This matches the large $H\ell$ limit of (\ref{eq:zero mode mass}),
as expected.

Note that the high energy limit (\ref{eq:high energy limit}) does
not distinguish between $H$ approaching infinity or $\ell$
approaching infinity.  Hence, it is possible to expand (\ref{eq:high
energy polynomial}) in the case of a flat stationary brane; i.e., $H
\rightarrow 0$ and $\ell \rightarrow \infty$.  This yields
\begin{equation}\label{eq:high energy flat space resonance}
    \beta^2 = i\rho^{-1}(\rho^2-\mu^2)(m^2-2\rho^2), \quad \rho =
    \omega H.
\end{equation}
Note that this result implicitly assumes the $m \rightarrow 0$
limit, since (\ref{eq:high energy polynomial}) is only valid when
$m\ell$ is non-infinite.  It is easy to confirm that this matches
the flat space resonance condition obtained in the low energy regime
(\ref{eq:flat space resonance}) when $m \rightarrow 0$.

\section{4-dimensional effective theory and the bound state mass
matrix}\label{sec:effective theory}

In this section, we show how many of the qualitative and
quantitative properties of the model derived from the spectral
analysis in Sec.~\ref{sec:spectral} are encapsulated by an
effective 4-dimensional action.  Here, $\phi$ and $q$ refer to the
fields in real space, rather than the spatial Fourier transforms.

To derive this action, let us write
\begin{equation}
    \phi(\tau,\mathbf{x},y) = \varphi_1(\tau,\mathbf{x})\xi(y), \quad
    q(\tau,\mathbf{x}) = \varphi_2(\tau,\mathbf{x}).
\end{equation}
We substitute these definitions into the full 5-dimensional action
(\ref{eq:action}) and explicitly perform the integrations over
$y$. We obtain the following 4-dimensional effective action:
\begin{equation}\label{eq:effective action}
    S_\eff = -\frac{1}{2} \int\limits_{\di\mathcal{M}_\b} \left[
    \sum_{i=1}^2
    ( \di^\alpha\varphi_i \di_\alpha \varphi_i
    + m_i^2 \varphi_i^2) + 2\betafour \varphi_1 \varphi_2
    \right].
\end{equation}
Here, Greek indices are raised and lowered with the brane metric
$h_{\alpha\beta}$.  We have defined the effective masses and
coupling as
\begin{gather}\nonumber
    m_1^2 = 2 m^2 \! \int_{y_\b}^\infty dy\,N^5(y)\xi^2(y)
    + 2 \int_{y_\b}^\infty dy\,N^3(y)
    [\xi'(y)]^2, \\ m_2^2 = \mu^2, \quad \betafour = \beta
    \xi(y_\b),\label{eq:effective theory def'ns}
\end{gather}
where $N(y) = H\ell(\sinh Hy)^{-1}$.  In deriving
(\ref{eq:effective action}), we have assumed
\begin{equation}\label{eq:effective theory assumption}
    \int_{y_\b}^\infty dy\,N^3(y)\xi^2(y) = \frac{1}{2}.
\end{equation}
This assumption is equivalent to demanding that the original bulk
field is normalisable under the Klein-Gordon inner product
(\ref{eq:KG inner product}).  Hence, our effective action only
gives a valid description of the bound state degrees of freedom in
the model.  The full action (\ref{eq:action}) is required if we
want to include continuum modes (i.e.,~scattering states) as well.

The 4-dimensional action (\ref{eq:effective action}) represents a
pair of coupled massive fields confined to the brane.  We can seek
mass eigenstate solutions (i.e.,~eigenfunctions of $\nabla^\alpha
\nabla_\alpha$) as follows:
\begin{subequations}
\begin{gather}
    \nabla^\alpha \nabla_\alpha \varphi_i = \rho^2 \varphi_i \,\,
    \Rightarrow \,\, \mathbf{M} \roarrow{\mathbf{\varphi}} =
    \rho^2 \roarrow{\mathbf{\varphi}},
    \\ \mathbf{M} = \left( \begin{array}{cc} m_1^2 & \betafour \\
    \betafour & m_2^2 \end{array} \right), \quad
    \roarrow{\mathbf{\varphi}} = \left( \begin{array}{c}
    \varphi_1 \\ \varphi_2 \end{array} \right) .
\end{gather}
\end{subequations}
This matrix equation has a non-trivial solution if $\rho^2$ is one
of the eigenvalues of the mass matrix $\mathbf{M}$, which are
\begin{equation}\label{eq:resonant masses}
    \rho^2_\pm = \frac{m_1^2 + m_2^2}{2} \left( 1 \pm \sqrt{1 + \frac{4(\betafour^2 - m_1^2 m_2^2)}
    {(m_1^2 + m_2^2)^2}} \right).
\end{equation}
These are the masses of the two normal modes of the system, which
should be identified with the Kaluza-Klein masses of bound state
resonances in the full 5-dimensional theory.

The fact that we can have at most two mass eigenstates in the
4-dimensional effective theory is consistent with the results of
Sec.~\ref{sec:bound states}, where we saw that there can be at
most two bound state resonances in the full 5-dimensional
treatment. Furthermore, we note that one of these modes becomes
tachyonic (i.e.,~$\rho^2_- < 0$) if
\begin{equation}\label{eq:effective theory critical coupling}
    \betafour^2 > \betafour^2_\text{crit} = m_1^2 m_2^2.
\end{equation}
This again matches our expectations: We have one and only one
tachyonic mode if the coupling is sufficiently large.

It is important to note that not all mass eigenstates in the
effective theory spectrum correspond to bound states of the full
5-dimensional model.  Recall that bound states must have
Kaluza-Klein mass within the mass gap, so any mass eigenstates
with $\rho^2_\pm > 9H/4$ are not true resonances.  We have that
$\rho_+^2$ is bounded from below by $\max(m_1^2,m_2^2)$, while
$\rho_-^2$ is bounded from above by $\min(m_1^2,m_2^2)$. Thus, a
necessary condition for there to be two bound state solutions in
the effective theory is
\begin{equation}
    m_1 < 3H/2, \quad m_2 = \mu < 3H/2.
\end{equation}
Qualitatively, this matches the results of Sec.~\ref{sec:bound
states}, where we saw that a necessary condition for the existence
of two bound states was that the field masses were not too heavy.
Unfortunately, one cannot translate these inequalities into more
stringent constraints on $m$ without knowing $\xi(y)$. We should
briefly mention the Minkowski brane case:  When $H=0$ there is no
mass gap and only tachyonic resonances with $\rho^2 < 0$ correspond
to bound states. Hence, in the effective theory there is only a
single tachyonic bound state for supercritical coupling, in
agreement with the conclusions of Ref.~\cite{Koyama:2005gh}.

We can quantitatively compare our effective theory to the results
previously obtained for the perturbed zero mode in
Sec.~\ref{sec:perturbed zero mode}.  We assume that the
extra-dimensional profile is the purely ``zero mode'' solution, as
given by Eq.~(\ref{eq:zero mode profile}):
\begin{equation}
    \xi(y) = \text{constant}.
\end{equation}
Under this assumption, the various integrals in (\ref{eq:effective
theory def'ns}) and (\ref{eq:effective theory assumption}) can be
reduced to elementary functions.  We find
\begin{gather}\nonumber
    \xi^2 = \ell^{-1}f(H\ell), \quad m_1^2 = m^2 g(H\ell), \\
    \betafour^2 = \ell^{-1}\beta^2 f(H\ell),\label{eq:effective theory identifications}
\end{gather}
where $f$ and $g$ are defined in Eqs.~(\ref{eq:f and g def}).  To
relate these back to the perturbed zero mode, recall that our
results for $\delta\rho_\text{zero}^2$ were derived for the case
when $m$ and $\beta$ were small.  This suggests that we expand the
effective theory resonant masses (\ref{eq:resonant masses}) under
the conditions $m_2^2 \gg \max(m_1^2,\betafour)$, which yields
\begin{equation}
    \rho_+^2 \sim m_2^2 + \betafour^2/m_2^2, \quad \rho_-^2 \sim m_1^2 -
    \betafour^2/m_2^2.
\end{equation}
Substituting (\ref{eq:effective theory identifications}) and $m_2 =
\mu$ into the above expression for $\rho_-^2$, we recover the the
5-dimensional prediction for $\delta\rho_\text{zero}^2$
(\ref{eq:zero mode mass}).  Finally, we note that the effective
theory prediction of the critical coupling (\ref{eq:effective theory
critical coupling}) in this regime is
\begin{equation}
    \betafour_\text{crit}^2 = \mu^2 m^2 g(H\ell) \sim \ell^{-1}
    f(H\ell) \betacrit^2,
\end{equation}
where we have made use of (\ref{eq:critical coupling approx}). The
relationship between $\betafour_\text{crit}^2$ and $\betacrit^2$
is entirely consistent with what we expect from (\ref{eq:effective
theory identifications}).

To summarize this section, we have shown that the bound state
sector of our model can be described by the effective action of a
pair of coupled scalars propagating on a de Sitter background. The
main complication is that the effective 4-dimensional masses and
coupling depend on higher-dimensional details, so moving from the
5-dimensional to 4-dimensional pictures requires the non-trivial
calculations performed in the rest of the paper.

\section{Numeric time domain analysis}\label{sec:numeric}

We now turn to the solution of the coupled equations (\ref{eq:EOMs
1}) subject to the boundary condition (\ref{eq:boundary 1}) in the
time domain.  As in Sec.~\ref{sec:spectral}, we will work with the
spatial Fourier transforms (\ref{eq:Fourier transform}) of $\phi$
and $q$ and omit the $\mathbf{k}$ subscript.

\subsection{Dimensionless wave equations in Poincar\'e coordinates}

In our numeric integrations, we elect to use dimensionless versions
of the Poincar\'e coordinates $t$ and $z$ (\ref{eq:coordinate
transform}), as well as the conformal time $\eta$ (\ref{eq:eta}). In
particular, we define
\begin{align}\nonumber
    \t & = {t}/{z_\star}, &
    \z & = {z}/{z_\star}, &
    \etahat & = {\eta}/{z_\star}, \\
    \dihat_t & = \di/\di\t, &
    \dihat_z & = \di/\di\z, &
    \dihat_\eta & = \di/\di\etahat,
\end{align}
where $z_\star$ is a length scale we specify later.  We rescale
the bulk field as follows
\begin{equation}\label{eq:psi_definition}
    \psi(t,z) = \ell^{1/2}\z^{-3/2}\phi(t,z),
\end{equation}
and write the brane trajectory as
\begin{equation}
    \t_\b = \gamma\etahat, \quad \z_\b = -H\ell\etahat, \quad d\etahat^2 = d\t_\b^2 - d\z_\b^2.
\end{equation}
Then, the wave equations (\ref{eq:EOMs 1}) and boundary condition
(\ref{eq:boundary 1}) reduce to
\begin{subequations}\label{eq:EOMs 4}
\begin{align}
    \label{eq:EOMs 4a} 0 & = [ \dihat_t^2 - \dihat_z^2 + V(\z) ] \psi , \\
    \label{eq:EOMs 4b} 0 & = [\dihat^2_\eta + A(\z_\b) \dihat_\eta + B(\z_\b) ] q
    + C(\z_\b) \psi_\b, \\ 0 & = ( \hat{\di}_n \psi )_\b - D(\z_\b)\psi_\b
    - E(\z_\b)q - F(\z_\b)\dihat_\eta q, \label{eq:EOMs 4c}
\end{align}
where the potential is
\begin{equation}
    V(\z) = k^2 z_\star^2+\tfrac{1}{4}(15+4m^2\ell^2)\z^{-2},
\end{equation}
the coefficients are
\begin{align}
    \nonumber A(\z_\b) & = 2H\ell\z_\b^{-1}, & D(\z_\b) & = -\tfrac{3}{2} \gamma \z_\b^{-1},\\
    \nonumber B(\z_\b) & = k^2z_\star^2 + \mu^2\ell^2 \z_\b^{-2} , & E(\z_\b) & = \tfrac{1}{2}\beta
    \ell^{3/2} \z_\b^{-5/2}, \\
    C(\z_\b) & = \beta \ell^{3/2} \z_\b^{-1/2}, & F(\z_\b) & = 0,
\end{align}
and the (flat-space) normal derivative operator is
\begin{equation}
    \dihat_n = -H\ell \, \dihat_t + \gamma \, \dihat_z.
\end{equation}
\end{subequations}
Note that even though the $F$ coefficient is zero for the problem
at hand, we retain it to keep the algorithm developed in the next
section reasonably general.

\subsection{The algorithm}

Previously, \citet{Seahra:2006tm} has developed a numeric
algorithm to solve for $\psi$ in equations similar to
(\ref{eq:EOMs 4}) when there is no brane-bulk coupling ($\beta =
0$) and the brane trajectory is arbitrary.  Here, we generalize
that procedure to the case when there is a dynamical degree of
freedom on the brane.

\begin{figure}
\begin{center}
\includegraphics{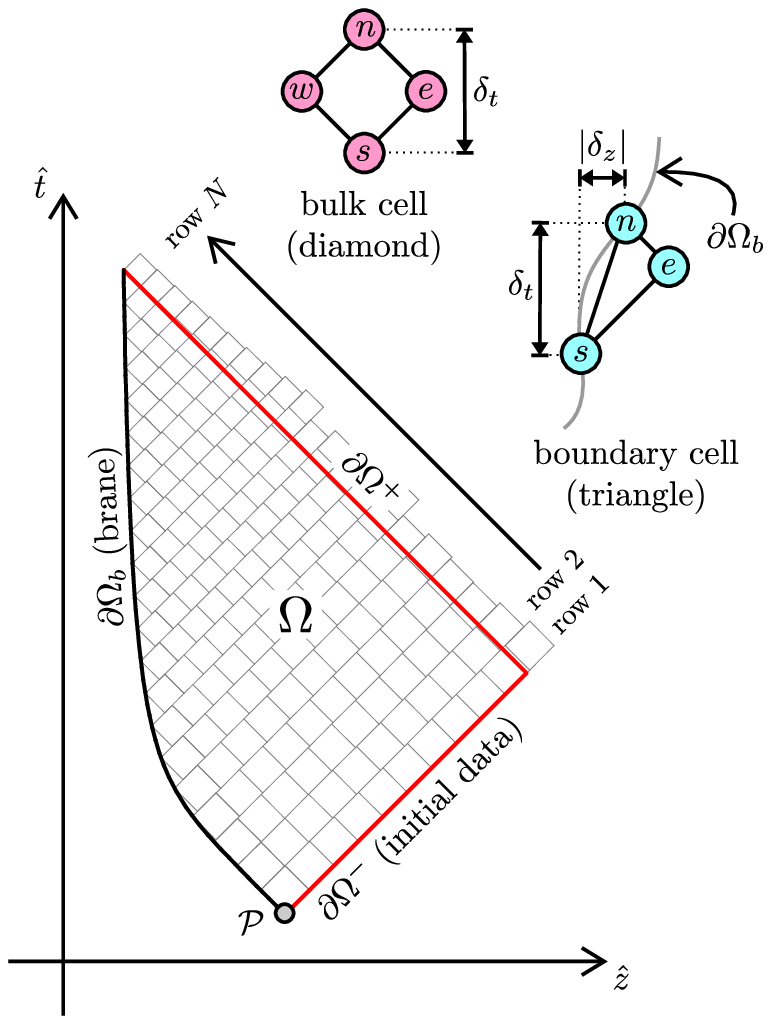}\\
\caption{The spacetime domain $\Omega$ over which we seek a
numeric solution of (\ref{eq:EOMs 4}). Superimposed on $\Omega$ is
a (particularly coarse) example of the computational grid we use
to discretize the problem.  Details of the two types of cellular
geometries are shown in the upper right corner.}\label{fig:grid}
\end{center}
\end{figure}

We seek the numeric solution for $\psi$ throughout a finite region
of the $(\t,\z)$ plane $\Omega$ depicted in Fig.~\ref{fig:grid},
as well as the solution for $q$ on a finite segment $\di\Omega_\b$
of the brane trajectory.  To have a well-posed Cauchy evolution
for the fields, we must specify initial data for $\psi$ on the
past null boundary of the computational domain $\di\Omega^-$, as
well as the value of $q$ and $\dihat_\eta q$ at $\mathcal{P} =
\di\Omega_\b \cap \di\Omega^-$.

As in Ref.~\cite{Seahra:2006tm}, our strategy will be to partition
$\Omega$ into a large number of small cells to construct a
discrete approximation to the fields.  In Fig.~\ref{fig:grid}, we
see that the cells bounding the brane take the shape of triangles
with one timelike and two null sides, while the cells interior to
$\Omega$ are diamonds with equal null sides.  By integrating the
bulk wave equation (\ref{eq:EOMs 4a}) across a diamond cell, one
can find out the value of $\psi$ at the northern node of the cell
(i.e., the uppermost corner) in terms of its value at the other
three nodes (cf.~\cite{Seahra:2006tm}):
\begin{equation}\label{eq:diamond evolution}
    \psi_n = -\psi_s + (\psi_w + \psi_e)(1-\tfrac{1}{8} \delta_t^2
    V_s) + \mathcal{O}(\delta_t^4),
\end{equation}
where $\delta_t = \t_n - \t_s$ is the difference between $\t$
evaluated at the top and bottom of the cell. The subscripts $n$,
$s$, $e$ and $w$ indicate the value of a quantity at the northern,
southern, eastern and western nodes, respectively.

The evolution across a triangular cell is more complicated due to
the presence of the boundary degree of freedom $q$.  It is useful to
re-write the second-order ODE (\ref{eq:EOMs 4b}) as a system of two
first-order equations
\begin{equation}
    \dihat_\eta q = p, \quad 0 = \dihat_\eta p + A(\z_\b)p +
    B(\z_\b)q + C(\z_\b) \psi_\b.
\end{equation}
These can be discretized across a given triangular cell using the
modified Euler method:
\begin{subequations}\label{eq:Euler}
\begin{eqnarray}
    q_n & = & q_s+\tfrac{1}{2} \delta_{\eta} (p_n+p_s)+\mathcal{O}(\delta_{\eta}^3),\\
    \nonumber
    p_n & = & p_s - \tfrac{1}{2} \delta_{\eta} ( A_n p_n + A_s p_s + B_n q_n +
    B_s q_s + \\ & & + C_n \psi_n + C_s \psi_s)+\mathcal{O}(\delta_{\eta}^3).
\end{eqnarray}
\end{subequations}
Also, by integrating (\ref{eq:EOMs 4a}) over the cell and making
use of (\ref{eq:EOMs 4c}) we obtain
\begin{multline}\label{eq:triangle}
    \psi_n = -\frac{6\delta_{\eta}(E_n q_n+E_s q_s+F_n p_n + F_s p_s)}
    {12+6\delta_{\eta}D_n+\delta_u\delta_vV_n} \\ - \frac{12+6\delta_{\eta}D_s+\delta_u\delta_vV_s}
    {12+6\delta_{\eta}D_n+\delta_u\delta_vV_n}\psi_s
    \\ + \frac{24-\delta_u\delta_vV_e}
    {12+6\delta_{\eta}D_n+\delta_u\delta_vV_n}\psi_e
    +\mathcal{O}(\delta_{t}^3).
\end{multline}
Here,
\begin{gather}
    \nonumber \delta_u = \delta_t - \delta_z, \quad \delta_v = \delta_t +
    \delta_z, \quad \delta_\eta = (\delta_t^2 - \delta_z^2)^{1/2},
    \\ \delta_t = \t_n - \t_s, \quad \delta_z = \z_n - \z_s.
\end{gather}
By solving the linear system (\ref{eq:Euler}) and
(\ref{eq:triangle}), we can obtain the future field values
$(q_n,p_n,\psi_n)$ in terms of the past field values
$(q_s,p_s,\psi_s,\psi_e)$.

Having written down these evolution formulae, the actual algorithm
for obtaining $\psi$ and $q$ is very similar to the one used in
Ref.~\cite{Seahra:2006tm}.  Consider the first row of cells in
Fig.~\ref{fig:grid}.  If we know the value of $\psi$ along
$\di\Omega^-$ as well as the value of $(q,p)$ at $\mathcal{P}$, then
application of the triangle and diamond evolution formulae in
sequence (i.e.~southwest to northeast) allows us to obtain
$(q,p,\psi)$ on the future boundary of the first row.  One can then
find the values of $\psi$ on the nodes of the past boundary of the
second row via interpolation. By repeating this procedure for rows 2
through $N$, one obtains the solution for the fields throughout
$\Omega$. As in Ref.~\cite{Seahra:2006tm}, we expect our numeric
solutions for $\psi$ and $q$ to converge like $\delta_t^2$ to the
actual field values.

\subsection{Results}\label{sec:results}

\subsubsection{Minkowski brane}

We now apply our algorithm to the problem introduced in
Sec.~\ref{sec:problem}.  First, let us consider the simpler case of
a brane stationary with respect to the Poincar\'e frame:
\begin{equation}
    H\ell = 0, \quad z_\b = \ell, \quad ds_\b^2 = -d\tau^2 + d\mathbf{x}^2;
\end{equation}
i.e., a brane with Minkowski geometry.  In this case, we select
\begin{equation}
    z_\star = \ell \,\, \Rightarrow \,\, \z_\b = 1.
\end{equation}
As discussed in Sec.~\ref{sec:low energy}, the time dependence of
any given resonant mode in this case is given by
\begin{equation}
    \phiR_\omega \propto e^{i\varpi t/\ell}, \quad \varpi = \sqrt{\rhohat^2+
    k^2\ell^2},
\end{equation}
where $\rhohat$ is a solution of the low energy resonance condition
(\ref{eq:low energy resonance}) with $H\ell = 0$.

As discussed in Sec.~\ref{sec:QNM}, we expect the late time
behaviour of $\phi$ and $q$ to be dominated by resonant mode
solutions at late times, no matter what initial data is selected on
$\di\Omega^-$.  Furthermore, if we wait long enough the resonant
mode with the smallest imaginary part --- which we call the
fundamental mode --- will dominate the other modes.  Hence for any
choice of initial data, we expect that
\begin{equation}\label{eq:template 1}
    q \xrightarrow[\infty]{\,\,\,t\,\,\,} \Re(\mathcal{C}_q e^{i\varpi_0 t/\ell}), \quad
    \phi_\b \xrightarrow[\infty]{\,\,\,t\,\,\,} \Re(\mathcal{C}_\phi e^{i\varpi_0
    t/\ell}),
\end{equation}
where $\mathcal{C}_q$ and $\mathcal{C}_\phi$ are complex numbers,
and $\varpi_0$ is the frequency of the fundamental mode.

\begin{figure}
\begin{center}
\includegraphics{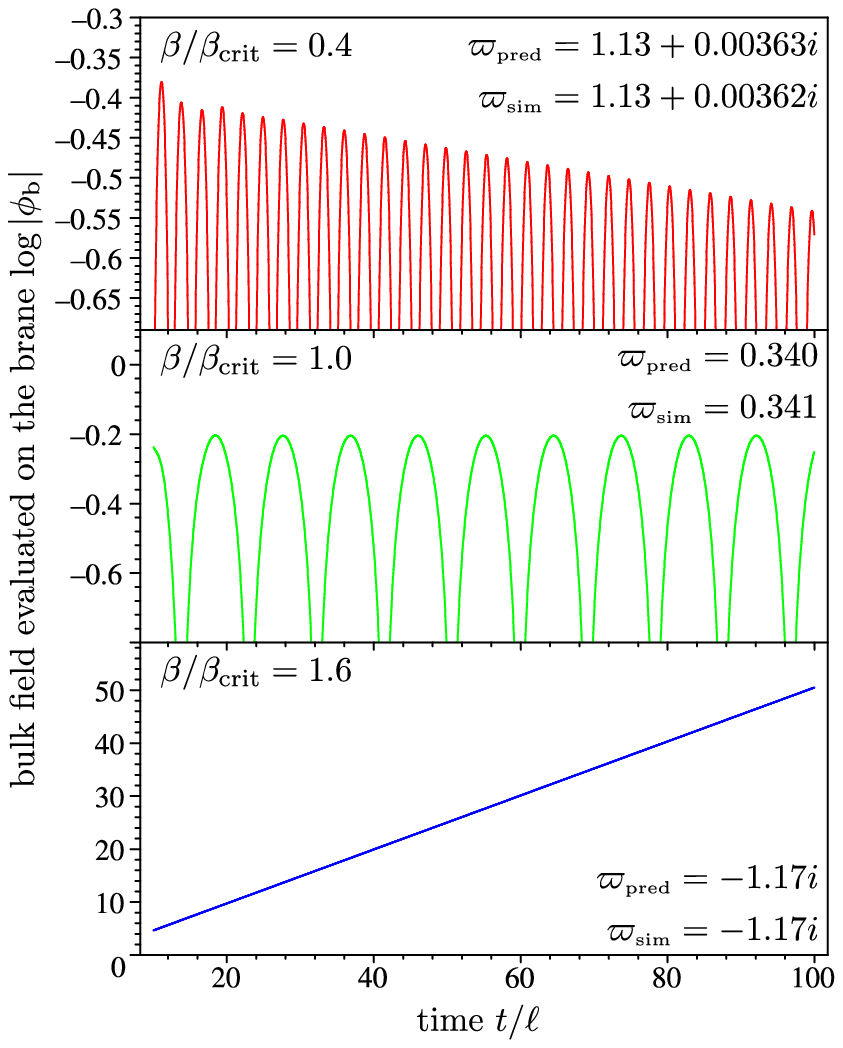}\\
\caption{Simulation results for a Minkowski brane with $\mu\ell =
1.23$, $m\ell = 2.37$ and $k\ell=0.340$ and the initial data
(\ref{eq:initial data}). The quoted values of
$\varpi_\text{\tiny{pred}}$ are predictions for the fundamental
frequency from the low energy resonance condition (\ref{eq:low
energy resonance}), while $\varpi_\text{\tiny{sim}}$ is the late
time frequency obtained by fitting the template (\ref{eq:template
1}) to the numeric waveforms.  This plot exhibits the three possible
classes of asymptotic field behavior when $H\ell = 0$.}
\label{fig:Minkowski waveforms}
\end{center}
\end{figure}
\begin{figure}
\begin{center}
\includegraphics{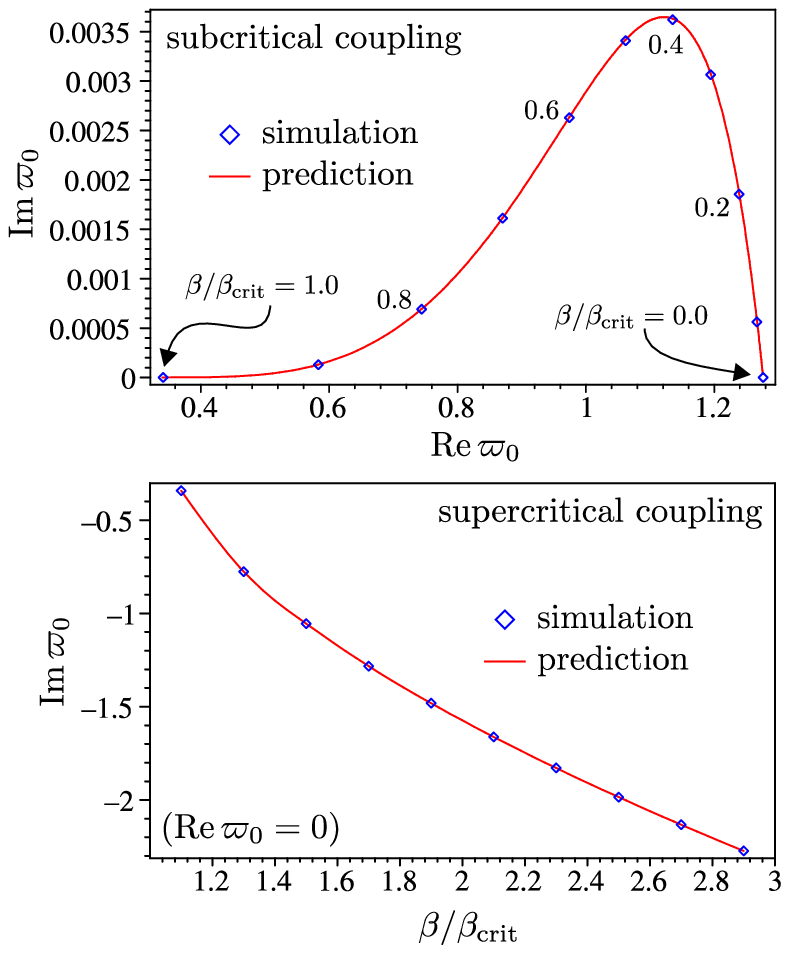}
\caption{Comparison between spectral analysis predictions for the
fundamental frequency and simulation results when $\mu\ell =
1.23$, $m\ell = 2.37$ and $k\ell=0.340$ (Minkowski
brane).}\label{fig:Minkowski compare}
\end{center}
\end{figure}

In Fig.~\ref{fig:Minkowski waveforms}, we show the results of our
numeric simulations for the following choice of initial data:
\begin{equation}\label{eq:initial data}
    q(\mathcal{P}) = 1, \quad p(\mathcal{P}) = 0 = \phi(\di\Omega^-).
\end{equation}
We find excellent qualitative and quantitative agreement between the
simulation results and the expectations for the fundamental
frequency from the spectral analysis.  In particular, we get
exponentially diverging fields (i.e.~tachyonic resonances) for large
coupling.  We further test the agreement between simulations and the
resonance condition (\ref{eq:low energy resonance}) in
Fig.~\ref{fig:Minkowski compare}.

\subsubsection{De Sitter brane}

For a de Sitter brane, we select $z_\star$ to be the position of the
brane when the mode under consideration exits the Hubble horizon,
which implies
\begin{equation}
    kz_\star = H\ell.
\end{equation}
Along the same lines as our discussion of the Minkowski brane above,
we expect that the late time behaviour of the fields in a numeric
simulation to be
\begin{subequations}\label{eq:template 2}
\begin{align}
    q & \xrightarrow[\infty]{\,\,\,\tau\,\,\,} \text{Re} \left[
    \mathcal{C}_q e^{(i\omega_0 - 3/2)H\tau} \right], \\
    \phi_\b & \xrightarrow[\infty]{\,\,\,\tau\,\,\,} \text{Re} \left[
    \mathcal{C}_\phi e^{(i\omega_0 - 3/2)H\tau} \right],
\end{align}
\end{subequations}
regardless of the particular choice of initial data.  Here, the
fundamental frequency $\omega_0$ is the solution of
(\ref{eq:resonance condition}) with the smallest imaginary part.

\begin{figure}
\begin{center}
\includegraphics{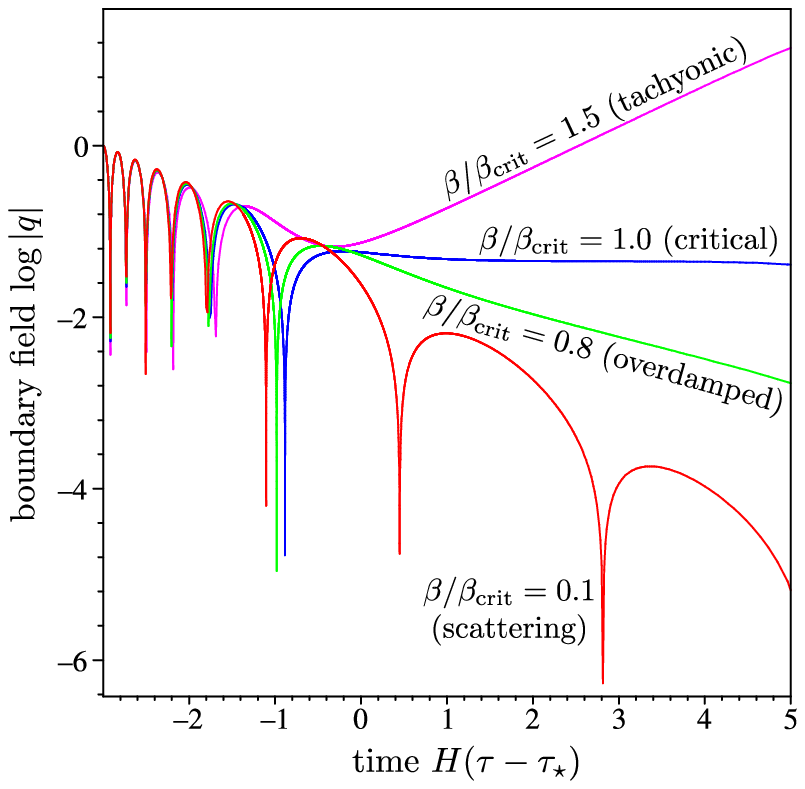}\\
\caption{Simulation results for a de Sitter brane with $\mu\ell =
0.40$, $m\ell = 2.1$ and $H\ell=0.20$ and the initial data
(\ref{eq:initial data}). Here, $\tau_\star$ is the proper time
coordinate on the brane when the mode exits the Hubble horizon.
Individual curves are labeled by the coupling parameter and the
classification of the fundamental mode solution of
(\ref{eq:resonance condition}) for these parameters---the four
possible varieties of late time behaviour when $H\ell > 0$ are all
seen in this plot.} \label{fig:deSitter waveforms}
\end{center}
\end{figure}
\begin{figure}
\begin{center}
\includegraphics{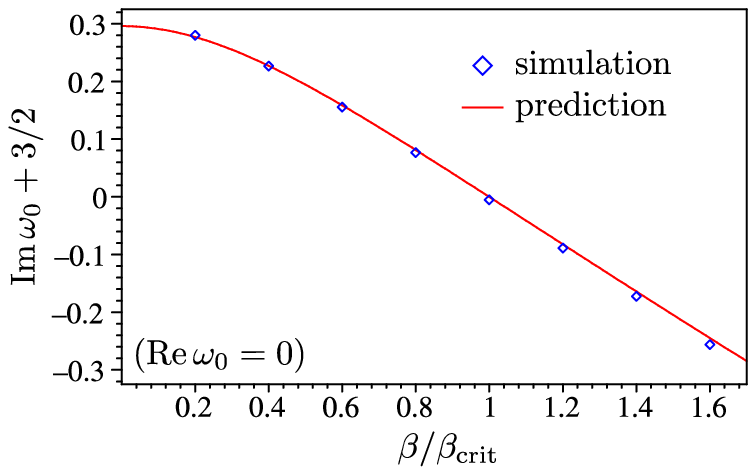}
\caption{Comparison between spectral analysis predictions for the
fundamental frequency from the full resonance condition
$\mathcal{R}_\omega = 0$ and simulation results when $\mu\ell =
1.00$, $m\ell = 2.24$ and $H\ell = 1.12$ (de Sitter brane). For this
choice of parameters, the fundamental resonance is never a
scattering state.}\label{fig:deSitter compare}
\end{center}
\end{figure}
In Fig.~\ref{fig:deSitter waveforms}, we show results for some of
our simulations.  The particular case considered is one in which
the fundamental mode is a scattering state for $\beta/\betacrit
\lesssim 0.65$, and an overdamped bound state for $0.65 \lesssim
\beta/\betacrit < 1$.  In Fig.~\ref{fig:deSitter compare}, we
quantitatively compare the value of the fundamental frequency
obtained from the spectral analysis and the value obtained from
simulations in a case where there is an overdamped bound state for
all subcritical couplings.  We find acceptable agreement between
the values of $\omega_0$ obtained in both approaches.

\section{Conclusions}\label{sec:conclusions}

In this paper we have studied resonant modes of a massive bulk field
in five-dimensional anti-de Sitter spacetime linearly coupled to a
massive field on a four-dimensional boundary de Sitter brane. Using
spectral scattering theory in 5 dimensions, we investigated the
resonant excitations of the system, identifying scattering states
which propagate to future null infinity and normalisable bound
states. By dimensionally reducing the full action, we have also
derived an effective 4-dimensional theory that gives a valid
description of the bound state resonances.  The effective theory is
essentially that of a pair of coupled massive brane-confined
scalars.

From the 5-dimensional treatment, we find that the bound state
sector of the spectrum is characterised by zero, one, or two
resonances. These can be divided into two categories: overdamped
and tachyonic modes which decay or diverge exponentially at late
times, respectively. We find that there is a critical coupling
$\betacrit$, above which there always exists a unique tachyonic
mode; conversely, no tachyonic bound state exists for subcritical
coupling.  The critical coupling is given analytically by
Eq.~(\ref{eq:critical coupling}), and we find that $\betacrit = 0$
if either field is massless.

All of these bound state properties are mirrored in the effective
4-dimensional theory: namely, the effective action predicts the
existence of at most two mass eigenstates, or normal modes, which
correspond to bound states of the higher-dimensional theory.
Furthermore, one of the normal modes is tachyonic when the coupling
exceeds a critical value, and that critical value is zero if either
of the effective field masses vanish.

One of the more interesting features of our model is the existence
of overdamped bound states, which do not appear in the case of a
Minkowski brane in an AdS bulk \cite{Koyama:2005gh}.  It is well
known that for a Minkowski brane, all bulk modes with real
Kaluza-Klein mass reduce to travelling waves far from the brane and
are hence non-normalisable.  However for a de Sitter brane, it is
possible to have normalisable bulk modes with Kaluza-Klein mass less
than $3H/2$ \cite{Garriga:1999bq}.  The overdamped bound states
found in this paper live within this mass gap, which explains why
they are not present when the brane is ``at rest'' ($H=0$). For
these stationary branes, one can only have a single tachyonic bound
state, and that state only exists for supercritical coupling.

The actual number of bound states for a given choice of parameters
is strongly influenced by the masses of the bulk and brane fields.
In particular, a necessary condition for the existence of two
bound state resonances is that both fields are ``light'' compared
to the Hubble scale.  In this context, the notions of light and
heavy refer to the behaviour of the individual fields at zero
coupling.  The brane scalar is considered to be light if it does
not oscillate on super-horizon scales when $\beta = 0$.
Quantitatively, this means that the boundary mass is less than
$3H/2$.  On the other hand, a light bulk scalar is one for which
there exists a normalisable bound state in the absence of any
brane-localised couplings.  We have numerically determined that
this is equivalent to demanding that the bulk mass squared $m^2$
is less than $\sim 4.50 \, H$ at low energy ($H\ell \ll 1$) and
less than $\sim 4.42 \, H$ at high energy ($H\ell \gg 1$).  We can
qualitatively recover these necessary conditions for multiple
bound states from the effective theory, but we can only obtain the
precise bound on $m$ quoted here from the full spectral formalism.

We studied the behaviour of the model's resonances in a number of
different limits in Sec.~\ref{sec:limiting cases}, and have
recovered several results previously available in the literature.
One case worth emphasizing involves the perturbation of the massless
de Sitter zero mode when $\beta$ and $m$ take on small values. Under
these conditions, we have demonstrated that there is complete
agreement between the predictions of the 5-dimensional spectral
formalism and the 4-dimensional effective theory for the
perturbative zero mode mass, critical coupling, etc.  In particular,
we find that the perturbed zero mode mass is given by
$\delta\rho^2_\text{zero} \sim
\tfrac{1}{2}m^2(1-\beta^2/\betacrit^2)$ for $H\ell \ll 1$ and
$\delta\rho^2_\text{zero} \sim
\tfrac{3}{5}m^2(1-\beta^2/\betacrit^2)$ for $H\ell \gg 1$. From the
condition that the the Kaluza-Klein masses of normalisable states
must lie within the mass gap, we have confirmed that a necessary
condition for a bulk bound state at zero coupling is $m^2/H^2 < 9/2$
for $H\ell \ll 1$, in agreement with the numerical results of
Sec.~\ref{sec:zero coupling}.

We have tested the predictions of the spectral theory against a new
numerical algorithm that is able to evolve the coupled brane and
bulk fields. We are able to confirm our spectral analysis
predictions for the fundamental frequencies by studying the
late-time behaviour for both a static (Minkowski) brane and a moving
de Sitter brane in the Poincar\'e coordinates. In particular we find
good quantitative agreement between the rate of decay or growth of
the boundary field and the imaginary part of the frequency of the
fundamental mode for both subcritical and supercritical coupling.

We have chosen to work with a moving brane Poincar\'e coordinates so
that the code can also be used to study cosmological branes with
arbitrary trajectories in AdS. Our numerical code can be used to
study the coupled evolution any system where the bulk and boundary
wave equations and boundary condition can be written in form given
in Eqs.~(\ref{eq:EOMs 4}). We will thus be able to study coupled
bulk and brane fields in situations where we have no analytic
results to compare against. For instance such a numerical code is
required to study the evolution of cosmological density
perturbations consistently coupled to bulk metric perturbations
during the matter- or radiation-dominated eras.

The simple model of linearly coupled bulk and brane fields on a de
Sitter brane which we have considered in this paper may be a
useful toy model for the study of inflaton perturbations linearly
coupled to scalar metric perturbations during inflation in a
higher-dimensional bulk. However the inflaton-metric interaction
involves derivative couplings which makes the problem considerably
more challenging. In particular we find that the bulk field is
coupled to an infinite tower of bulk modes
\cite{Koyama:2004ap,Koyama:2005ek,Koyama:2007as}. We will return
to this problem in future work.

\begin{acknowledgments}
AC is supported by FCT (Portugal) PhD fellowship
SFRH/BD/19853/2004.  KK and SSS are supported by PPARC. AM is
supported by PPARC grant PPA/G/S/2002/00576.
\end{acknowledgments}

\bibliography{deSitter_coupled}

\end{document}